\documentclass[aps,pre,reprint,superscriptaddress,amsmath,amssymb,showpacs,floatfix]{revtex4-1}
\usepackage{graphicx}
\usepackage{bm}
\usepackage{color}
\usepackage{soul}
\usepackage{tikz}
\newcommand*\circled[1]{\tikz[baseline=(char.base)]{
            \node[shape=circle,draw,inner sep=0.3pt] (char) {#1};}}

\begin{document}
\title{ Quantumness and thermodynamic uncertainty relation of the finite-time Otto cycle }

\author{Sangyun Lee}
\affiliation{Department of Physics, Korea Advanced Institute of Science and Technology, Daejeon 34051, Korea }

\author{Meesoon Ha}
\email[]{msha@chosun.ac.kr}
\affiliation{Department of Physics Education, Chosun University, Gwangju 61452, Korea}

\author{Hawoong Jeong}
\email[]{hjeong@kaist.edu}
\affiliation{Department of Physics and Institute for the BioCentury, Korea Advanced Institute of Science and Technology, Daejeon 34141, Korea}


\date{\today}

\begin{abstract}
To reveal the role of the quantumness in the Otto cycle and to discuss the validity of the thermodynamic uncertainty relation (TUR) in the cycle, we study the quantum Otto cycle and its classical counterpart. In particular, we calculate exactly the mean values and relative error of thermodynamic quantities. In the quasistatic limit, quantumness reduces the productivity and precision of the Otto cycle compared to that in the absence of quantumness, whereas in the finite-time mode, it can increase the cycle's productivity and precision. Interestingly, as the strength (heat conductance) between the system and the bath increases, the precision of the quantum Otto cycle overtakes that of the classical one. Testing the conventional TUR of the Otto cycle, in the region where the entropy production is large enough, we find a tighter bound than that of the conventional TUR. However, in the finite-time mode, both quantum and classical Otto cycles violate the conventional TUR in the region where the entropy production is small. This implies that another modified TUR is required to cover the finite-time Otto cycle. Finally, we discuss the possible origin of this violation in terms of the uncertainty products of the thermodynamic quantities and the relative error near resonance conditions.
\end{abstract}

\maketitle
 
\section{introduction}
\label{intro}
Understanding quantum effects on thermodynamics is one of the main goals of quantum thermodynamics. Quantum thermodynamic machines are cornerstones of related research that provide comprehensive understanding and important applications. Therefore, it is worthwhile to reveal the quantum effects on physical systems by direct comparisons of quantum and classical thermodynamic machines.

To date, various quantum devices have been realized, which give many lessons. Among them, finite-level quantum devices were realized with spin~\cite{deAssis2019PRL-QOHE-experiment, peterson2018experimental} and quantum dots~\cite{josefsson2018quantum}, but their classical counterparts are still questionable. In that sense, a harmonic oscillator is a good choice since it is an infinite-level system and has a natural classical counterpart. In addition, the harmonic oscillator system can be realized with an ion trap as a cyclic-quantum-heat engine~\cite{Rossnagel2016single}. In particular to the Otto cycle, the closed form of a density matrix can be earned even in the finite-time mode~\cite{Insinga2016Thermodynamical}. Due to this benefit, basic properties including the efficiency and power of the Otto cycle have been calculated from a one-time energy function in a couple of recent studies~\cite{Insinga2018quantum, Lee2020Finite}. 

In nonequilibrium physics, fluctuations of the thermodynamic quantities are major targets to be measured and studied. A few decades ago, symmetry in the fluctuations was found, which was named, {\em the fluctuation theorem} (FT)~\cite{Seifert2012Stochastic}. The FT enables us to measure the free energy of a nonequilibrium process, and it helps us find other forms of the thermodynamic second law for various systems, including the information engine~\cite{Koski2014Experimental}. Recently, the bound of the fluctuations has gathered much attention because the thermodynamic uncertainty relation (TUR) states that the relative fluctuations (relative errors) of thermodynamic quantities are bounded by a value inversely proportional to the entropy production of the process. This means that the TUR governs the trade-off between the relative error and entropy production, and it implies that to reduce the relative error, we have to pay a thermodynamic cost, i.e., entropy production. The TUR bound was proven for classical Markov jump processes with even-parity state variables~\cite{Barato2015Thermodynamic, Gingrich2016Dissipation}. However, the validity of the TUR bound for more complex systems, such as underdamped Langevin~\cite{Van2019uncertainty, Chun2019Effect, Lee2019Thermodynamic, Fischer2020Free} and quantum systems~\cite{Timpanaro2019thermodynamic, Guarnieri2019thermodynamics, Carollo2019unraveling, miller2020thermodynamic}, is still questionable. Many researchers have therefore tried to find a bound that is more broadly applicable and tight enough to give fruitful insight into a wider range of nonequilibrium systems~\cite{Hasegawa2019Fluctuation, Horowitz2020Thermodynamic, Vu2020Generalized, Hasegawa2020quantum}. 

In this paper, we study two types of the finite-time Otto cycle with a harmonic oscillator. 
One is a quantum Otto cycle, and the other is a classical Otto cycle.  
To reveal the quantum effects on the Otto cycle, we calculate exactly the mean values and fluctuations of a number of thermodynamic quantities, such as work, hot heat, and cold heat, for both cases. In the case of the quantum Otto cycle, we measure operational work and heat~\cite{Breuer2006TheTheory}. Based on our results, we find that quantum relative errors are bounded by classical relative errors in the quasistatic limit,
and some counter-intuitive result arises that quantum effects increase the precision of the quantum cycle in the finite-time mode. Testing the validity of the conventional TUR for both Otto cycles,  in the region where the entropy production is large enough, the Otto cycles exhibits a tighter bound than that of the conventional TUR. Unlike the quasistatic limit, however, in the finite-time mode, the violation of the conventional TUR is observed in both Otto cycles near the resonance conditions when the entropy production is small. We discuss the precision of the finite-time Otto cycle and the violation of the conventional TUR, and we argue their possible origins with some intuitive explanations.

The rest of this paper is organized as follows. In Sec.~\ref{system}, we describe a finite-time Otto cycle with a harmonic oscillator, and we explain how to treat two different types of heat baths, i.e., quantum and classical. We also show how to calculate the correlation functions. In Sec.~\ref{result}, we present the quasistatic results of both Otto cycles, including the uncertainty products of the thermodynamic quantities and the test of the conventional TUR. Finally, in Sec.~\ref{summary}, we conclude this paper with a summary and some open questions.

\begin{figure}[]
 \includegraphics[width=\columnwidth]{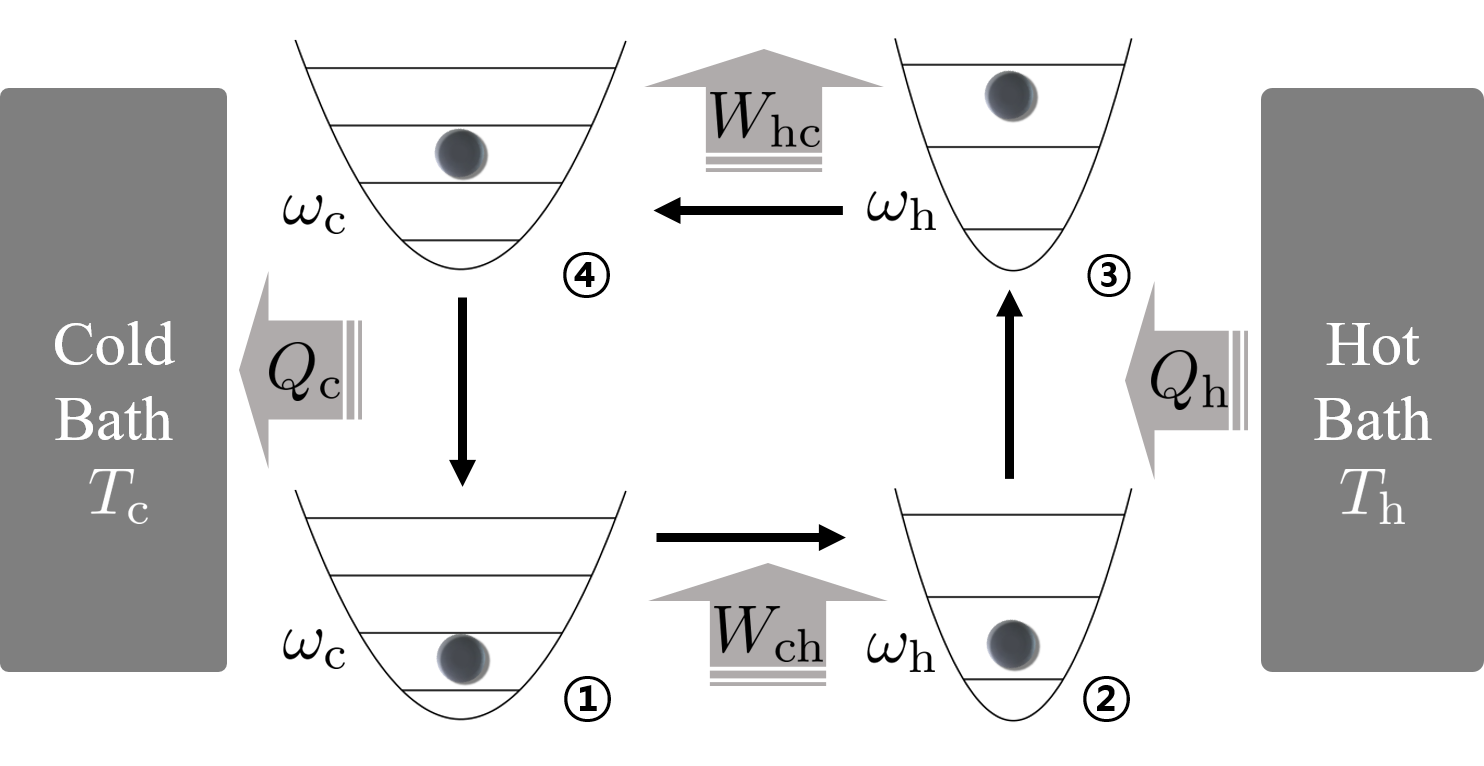}
\caption{ An Otto cycle is schematically illustrated. It consists of four thermodynamic processes with a harmonic oscillator, a hot heat bath of temperature $T_{\rm h}$ and a cold bath of $T_{\rm c}$:  The black solid right-arrow ($\rightarrow$) represents an adiabatic compression with work $W_{\rm ch}$ and the black up-arrow ($\uparrow$) represents a hot isochore with heat $Q_{\rm h}$. Correspondingly, the black solid left-arrow ($\leftarrow$) represents an adiabatic expansion with $W_{\rm hc}$ and the black down-arrow($\downarrow$) represents a cold isochore with $Q_{\rm c}$. Here $\omega_{\rm c}$ and $\omega_{\rm h}$ represent the harmonic frequencies at the cold isochore and  the hot isochore, respectively.}
\label{fig:fig1}
\end{figure}

\section{system}
\label{system}

\subsection{Working fluid and heat bath}
For a finite-time Otto cycle as shown in Fig.~\ref{fig:fig1}, we choose a harmonic oscillator as a working fluid, which is written as 
\begin{align}
    \hat{H}(t) =\frac{ \hat{p}^2 }{2m} + \frac{m\omega^2(t)\hat{x}^2}{2}
    = \frac{\hbar\omega(t)}{2} (\hat a^{\dagger} \hat a + 1)
    \label{eq:QHO}
,\end{align}
where $m$ is mass, $\omega(t)$ is the harmonic frequency of the working fluid, the hat notation ($\hat \cdot$) represents an operator, $\hat a = \sqrt{\frac{m\omega}{2\hbar}} \hat x + \frac{i}{\sqrt{2m\hbar\omega}} \hat p $ is an annihilation operator, and $\hat a^{\dagger} = \sqrt{\frac{m\omega}{2\hbar}} \hat x - \frac{i}{\sqrt{2m\hbar\omega}} \hat p $ is a creation operator. The inverse of $\omega(t)$ accords with the volume of the working fluid~\cite{Romero-Roch2005Equation}. For example, when we compress the volume of an isolated harmonic oscillator (increase $\omega$), the effective volume ($\sqrt{\langle \hat{x}^2\rangle }$) decreases and the energy of the working fluid increases. As the volume of the harmonic oscillator is changed, the corresponding eigenstates~$|n\rangle = \frac{(\hat a^\dagger)^n}{\sqrt{n!}}| 0 \rangle$ are changed, so that the coherence of the working fluid~\eqref{eq:QHO} can be generated in finite-time mode. By replacing operators ($\hat x$, $\hat p$) with state variables ($x$, $p$), we get a classical working fluid, which cannot have the coherence intrinsically, and we investigate the role of the coherence from a comparison between the quantum Otto cycle and its classical counterpart in Sec.~\ref{result}. 

For the isochoric processes of the quantum Otto cycle, the dynamics of the density matrix~$\hat{\rho} (t)$ is governed by the Lindblad equation with the superoperator $\mathcal{L} $~\cite{Breuer2006TheTheory} 
\begin{align}\label{eq:Qmas}
    \frac{{\rm d} \hat{ \rho } (t) }{{\rm d} t} =& 
    -\frac{i}{\hbar}[ \hat H(t), \hat\rho(t)] + \mathcal{L}(\hat{\rho}(t))
\end{align}
where 
\begin{align}\label{eq:SuperOp}
   {\mathcal{L}}(\hat\rho(t)) =& 
     \frac{\gamma}{2} ( \bar{n} + 1 ) [\hat a \hat \rho (t) \hat a^{\dagger}- \frac{1}{2} (\hat a^{\dagger} \hat a    \hat\rho (t) + \hat \rho (t) \hat a^{\dagger}\hat a )  ]\\
    &+ \frac{\gamma}{2} \bar{n}  [\hat a^{\dagger}\hat \rho (t) \hat a - \frac{1}{2} (\hat a\hat a^{\dagger} \hat   \rho(t) +\hat \rho(t)\hat a\hat a^{\dagger} ) ]
.\end{align} 
Here, $\gamma$ is the heat conductance of the heat bath, $\bar n = 1/(e^{\frac{\hbar\omega}{k_{B}T}}-1)$ is the expected value of the number operator of the heat bath. In the right-hand side of Eq.~\eqref{eq:SuperOp}, the first line decreases the energy of the working fluid, while the second line increases the energy of the working fluid. 

For the isochoric processes of the classical Otto cycle, the Fokker--Planck (FP) equation governs the dynamics of the probability density~$\rho(x,p,t)$. With $\vec q = ( x,  p)$, the FP equation is written as 
\begin{align}
    \partial_{t} \rho(x,p,t) = 
    -\vec\nabla_{q}
    \cdot
    [\mathcal{A}
    \cdot \vec{q}
    -
    \mathcal{B}
    \cdot
    \vec\nabla_{q}]
    \rho(x,p,t),
\label{eq:FP}\end{align}
where
\begin{align}
    \mathcal{A} = \left(\begin{smallmatrix}
    0 &  \frac{1}{m}  \\
    -m \omega^2(t) & -\frac{\gamma}{4}
    \end{smallmatrix}\right); \quad
    \mathcal{B} =  \left(\begin{smallmatrix}
    0 & 0             \\
    0                & \frac{m \gamma  }{4} T
    \end{smallmatrix}\right)
.\end{align} 
We can rewrite Eqs.~\eqref{eq:Qmas} and~\eqref{eq:FP} in Langevin forms, where the difference between the structures of the two baths is clearly revealed. For the quantum case, the Langevin form is called the quasiclassical Langevin equation~\cite{gardiner2004quantum}, and it is written as 
\begin{align}
        \partial_{t} x &= \frac{p}{m} - \frac{\gamma}{4} x 
        + \sqrt{ \frac{\gamma k_{B} \tilde T  }{4 m\omega^2 }  } \eta_{x}(t),   \\
    \partial_{t} p &= -m\omega^2 x 
        - \frac{\gamma}{4} p + \sqrt{\frac{ m\gamma k_{B} \tilde T }{4}} \eta_{p}(t)
\label{eq:QCL}\end{align}
where $\eta_{x}$ and $\eta_{p}$ represent two independent Gaussian white noises with $\langle \eta_i (t) \rangle = 0$ and $\langle \eta_{i}(t) \eta_{j}(t')  \rangle = 2\delta_{ij} \delta(t-t')$, respectively, and $\tilde T$ denotes the effective temperature of the quantum heat bath, which is $\hbar \omega (\bar n + 1/2)/k_{B}$. 

The Langevin equation for the classical heat bath is written as 
\begin{align}
    \partial_{t} x &= \frac{p}{m} ,  \\
        \partial_{t} p &= -m\omega^2 x 
        - \frac{\gamma}{4} p + \sqrt{\frac{ m\gamma  k_{B} T }{4}} \eta_{p}(t).
    \label{eq:CL}
\end{align}
On the right-hand side of the first line of Eq.~\eqref{eq:QCL}, there is a positional thermostat that cannot exist in classical physics, Eq.~\eqref{eq:CL}. This additional thermostat originates from the condition of a completely positive map, which ensures the positive definiteness of density matrix $\hat \rho(t)$ for any Hamiltonian~\cite{Breuer2006TheTheory}. 
Since work is either extracted or exerted through the potential that depends on the position, the positional thermostat leads to interesting differences between the two Otto cycles in the short-time limit, which are discussed in Set.~\ref{result}.

\subsection{Otto cycle}

The Otto cycle consists of two adiabatic and two isochoric processes, as illustrated in Fig.~\ref{fig:fig1}. In an adiabatic process, a heat bath is detached from a working fluid and the volume $\omega^{-1}(t)$ is changed. In an isochoric process, a heat bath is attached to a working fluid and the volume is fixed. The dynamics of quantum and classical working fluid is governed by Eqs.~\eqref{eq:Qmas} and~\eqref{eq:FP}. Sequentially, one pass of the Otto cycle can be written as follows:
\begin{enumerate}
	\item Adiabatic compression ($\circled{\bf 1} \to \circled{ \bf 2}$)
	\item Hot isochore ($\circled{\bf2}\to \circled{\bf3}$)
	\item Adiabatic expansion ($\circled{\bf3} \to\circled{\bf 4}$)
	\item Cold isochore  ($\circled{\bf4} \to\circled{\bf 1}$)
\end{enumerate}
In the adiabatic compression process, without a heat bath, the volume of the working fluid is compressed, $\omega^{-1}_{\rm c}\rightarrow \omega^{-1}_{\rm h}$, and work is exerted on the working fluid. In the hot isochore, the working fluid contacts a hot heat bath $T_{\rm h }$ at fixed volume $\omega^{-1}_{\rm h}$, and heat is exchanged between the working fluid and the hot heat bath. In an adiabatic expansion process, the working fluid is disconnected from the heat bath, and the volume of the working fluid expands, $\omega^{-1}_{\rm h} \rightarrow \omega^{-1}_{\rm c}$, by which work is extracted. At last, in the cold isochore, the working fluid contacts a cold heat bath and heat is again exchanged between the working fluid and the cold heat bath.

For the Otto cycle, we denote thermodynamic quantities as follows:
\begin{align}
    \hat W =& \hat W_{\rm hc} - \hat W_{\rm ch}\nonumber\\
    =&\hat H(t_3) -\hat H(t_4)  - \hat H(t_2) + \hat H(t_1) \\
    \hat Q_{\rm h} =& \hat H(t_3) - \hat H(t_2) \\ 
    \hat Q_{\rm c} =& \hat H(t_4) - \hat H(t_5) \\
    \hat \Sigma =& - \hat Q_{\rm h} /T_{\rm h} + \hat Q_{\rm c} /T_{\rm c}
    \label{eq:EP}
.\end{align}
Here $t_1$ and $t_3$ is the start time of the adiabatic compression and expansion process, $t_2$ ($t_4$) is the start time of the cold (hot) isochore, and $t_5 $ is the end time of the Otto cycle. It is noted that the fluctuations of thermodynamic quantities depend on the start point of the Otto cycle. In this paper, we set the compression step as the start time of the Otto cycle, and we focus on the cyclic steady states of the Otto cycle. We also point out that the last term in Eq.~\eqref{eq:EP} is the Clausius entropy of the Otto cycle, i.e., the entropy production of the heat bath.

Depending on the model parameters, the Otto cycle works as an engine, a refrigerator, or a heater. When $\langle \hat W \rangle > 0 $, $ \langle \hat Q_{\rm h} \rangle > 0 $ and $ \langle \hat Q_{\rm c} \rangle > 0 $, it works as an engine and transforms the partial heat flow from the hot bath to the cold bath into work. When $\langle \hat W \rangle < 0 $, $ \langle \hat Q_{\rm h} \rangle < 0 $, and $ \langle \hat Q_{\rm c} \rangle > 0 $, it works as a refrigerator and cools the cold bath by consuming work. In the other case, the Otto cycle transforms work into heat and heats the cold bath or hot bath, which corresponds to a heater, also called a useless machine~\cite{Lee2020Finite}. 

Other interesting quantities are the relative fluctuations, which can be used as the measure for the precision of a thermodynamic machine. 
We denote relative errors, $ \epsilon_{\hat A} \equiv \langle \Delta \hat A^2 \rangle / \langle \hat A \rangle^2 $ as the precision measure of the Otto cycle. Calculating fluctuations, the two-time correlation functions of the Hamiltonian are required, {\em e.g.} the second moment of work is written as 
\begin{align}
    \langle \hat W^2 \rangle
    =& \sum_{i=1}^{4}\langle \hat H ^2 (t_i) \rangle  
    + \sum_{\substack{i, j = 1\\ i \neq j }}^{4} (-1)^{i+j} \langle \hat H (t_i) \hat H (t_j) \rangle_{\rm s}  
    \label{eq:work2nd}
\end{align}
where \begin{align}
\langle \hat  A(t )\hat B(t') \rangle_{\rm s} \equiv \frac{1}{2}\langle \hat A(t) \hat B(t') + \hat B(t')\hat A(t) \rangle.
\label{eq:s-average}
\end{align} 
The relations between Hamiltonian moments and thermodynamic quantities are the same even in the classical Otto cycle. To yield meaningful physical quantities including relative error, calculations of both one-time and two-time moments of energy are required.

\subsection{ Calculation of correlation functions}

Correlation functions are basic blocks to calculate physical quantities such as work, heat, and fluctuations. The quasistatic limit yields a simple joint probability for energy, with which the correlation functions can be calculated. In the adiabatic process, the number of quanta ($\hat n = \hat a^\dagger \hat a$) is conserved for each sample and the initial and final states are fully correlated. On the other hand, in the isochoric process, all information dissipates and the correlation between the initial and final states vanishes.
Accordingly, in the quasistatic limit, the joint probability for the quantum Otto cycle is written as 
\begin{align}
\label{eq:joint}
    p^{\rm cyc}(n_1,n_2,n_3,n_4,n'_1) =& 
    p^{\rm \bf Q}_{\rm c} (n_1)  \delta_{n_1, n_2} p^{\rm \bf Q}_{\rm h} (n_3)\\
    &\times \delta_{n_3, n_4}  p^{\rm \bf  Q}_{\rm c} (n'_1),\nonumber
\end{align}
where {\bf Q} stands for the quantum heat bath, 
\begin{align}
p^{\rm \bf Q}_{\rm j} (n) \equiv \frac{e^{-\beta_{\rm j} \hbar\omega_{\rm j} (n + 1/2 )}}{Z_{\rm j}}
\end{align}
 with ${\rm j}\in\{\rm c, h\}$, $\beta_{\rm j}=\frac{1}{k_{\rm B}T_{\rm j}}$, and
$Z_{\rm j} = 1/[e^{\frac{\beta_{\rm j}\hbar\omega_{\rm j}}{2}} - e^{-\frac{\beta_{\rm j}\hbar\omega_{\rm j}}{2}}],$ 
and $\delta_{n{_\ell}, n{_{\ell'}}}$ is the Kroneck delta that is 1 if $n_{\ell}=n_{\ell'}$; 0 otherwise.

%
%
For the case of a classical system, an action 
\begin{align}
I = \frac{1}{2\pi}\oint p {\rm d}x
\end{align} is an adiabatic invariant.  For a harmonic oscillator with energy $E$, the action is $\frac{4 E}{ \omega \pi}\int^{1}_{0} (1-y^{2})^{1/2} dy$. From the adiabatic invariant the final energy is yielded as $E_{\rm f} = E_{\rm i} \omega_{\rm f}/\omega_{\rm i}$, where $\rm i$ stands for the initial state and $f$ for the final state. 
For a harmonic oscillator with energy $E$, the action is $\frac{4 E}{ \omega \pi }\int^{1}_{0} (1-y^{2})^{1/2} dy$. From the adiabatic invariant, the final energy is given as $E_{\rm f} = E_{\rm i} \omega_{\rm f }/\omega_{\rm i}$. 
Thus, the joint probability, $\rho^{\rm cyc}$, is written as  
\begin{align}
    \rho^{\rm cyc}( E_1, E_2, E_3, E_4 , E'_1)
    =& \rho^{\rm \bf C}_{\rm c} (E_1) \delta(E_2 - E_1 \frac{\omega_{\rm h}}{\omega_{\rm c}} ) \\
    &\times \rho^{\rm \bf C}_{\rm h} (E_3)  \delta(E_4 - E_3 \frac{\omega_{\rm c}}{\omega_{\rm h}})
    \rho^{\rm \bf C}_{\rm c} (E'_1)
    \nonumber
\label{eq:cyclejointprob}
\end{align}
where {\bf C} stands for the classical bath, 
\begin{align}
\rho^{\rm \bf C}_{\rm j} (E) \equiv {e^{-\beta_{\rm j} E }}/\beta_{\rm j}^{-1}.
\end{align}
If there is an isochore between the initial and final state, then the initial and final energy are uncorrelated. In terms of initial energy $E_{\rm i}$ and final energy $E_{\rm f}$, the relation is written as $\langle E_{\rm i} E_{\rm f} \rangle = \langle E_{\rm i} \rangle \langle E_{\rm f} \rangle$.
For the other case, if only an adiabatic process between the initial and final states exists, then the final density matrix (function) is the same as the initial density matrix (function).
In other words, $\langle E_{\rm i} E_{\rm f} \rangle = \frac{\omega_{\rm f}}{\omega_{\rm i}}\langle E_{\rm i}^2 \rangle$. 
These two relations are true for both quantum ({\bf Q)} and classical ({\bf C}) Otto cycles in the quasistatic limit. 

In the finite-time mode, we calculate the correlation functions from their governing equation. For both quantum and classical Otto cycles, cyclic steady states have a Gaussian form due to the structures of Eq.~\eqref{eq:Qmas} and Eq.~\eqref{eq:FP}. When the form of the state is Gaussian, the first and second moments of the corresponding random variables contain all information of the states. 
The Otto cycle has the left-right symmetry, so that the first moments of $x$ and $p$ are zero. Thus, the combinations of second moments,
\begin{align}
\hat H(t) =&\hat p^2/2m + m\omega^2(t) \hat{x}^2 / 2,\\
\hat{L}(t) =&\hat p^2/2m - m\omega^2(t) \hat{x}^2 / 2,\\
\hat{D}(t) =& \omega(t) (\hat{x}\hat{p} + \hat{p}\hat{x}) /2,     
\end{align}
are sufficient to describe the cycle~\cite{Kosloff2014-Review}. 
Since $\langle \hat L(t) \rangle $ and $\langle \hat D(t) \rangle$ are non-zero only when the off-diagonal components of the density matrix are non-zero, these components can be used to measure coherence~\cite{Kosloff2014-Review}.
The dynamics of these vectors are governed by a linear equation.
\begin{align}
    \frac{\rm d}{{\rm d}t}
    \vec\phi^{\rm k}(t)
    =
    \mathcal{M}^{\rm k}_{\rm j}
    \vec\phi^{\rm k}(t),
\label{eq:adiabaticHLDI}
\end{align}
where the subscript $\rm j\in\{\bf a, i\}$ for the index of the process, either adiabatic or isochoric, and the superscript $\rm k\in\{\bf Q, C\}$ for the treatment type of heat bath with
\begin{align}
    \vec\phi^{\rm \bf Q}(t) =& (\langle \hat H(t)\rangle,\langle \hat L(t) \rangle,\langle \hat D(t)\rangle,\langle \hat I\rangle),\\
    \vec\phi^{\rm \bf C}(t) =& (\langle  H(t)\rangle,\langle  L(t) \rangle,\langle  D(t)\rangle, 1)
.\end{align}
Here, $\hat I$ is an identity operator. 
In the adiabatic process, a quantum vector $\vec\phi^{\rm \bf Q}(t)$ and a classical vector $\vec\phi^{\rm \bf C}(t)$ are governed by the same matrix $\mathcal M^{\rm \bf C}_{\bf a}=\mathcal M^{\rm \bf Q}_{\bf a}=\mathcal M_{\bf a}$, which is written as 
\begin{align}
    \mathcal{M}_{\bf a} 
    =
    \omega(t)
    \begin{pmatrix}
    \frac{\dot \omega(t) }{\omega^2(t)} & - \frac{\dot{\omega}(t)}{\omega^2(t)} & 0 & 0  \\
    - \frac{\dot \omega(t) }{\omega^2(t)} &  \frac{\dot{\omega}(t)}{\omega^2(t)} & - 2 & 0  \\
    0 & 2 & \frac{\dot\omega(t)}{\omega^2(t)} & 0 \\
    0 & 0 & 0 & 0
    \end{pmatrix},
\label{eq:adia_mat}
\end{align}
This is due to the properties of a harmonic oscillator based on a consequence of Ehrenfest's theorem, and it is valid solely for the quadratic potential because $V'(x)=\frac{{\rm d}V(x)}{{\rm d}x}$ is linear in $x$, and thus, $V(\langle x\rangle)=\langle V(x)\rangle$. Here $\bf a\in \{\rm com,\, \rm exp \}$  denotes either the adiabatic compression or the adiabatic expansion process.

In the adiabatic process, Hamiltonian $\hat H(t)$, Lagrangian $\hat L(t)$, and correlation $\hat D(t)$ are coupled. When the time protocol $0\leq t \leq \tau$ is given as $\omega(t)=1/[ \omega_i^{-1}+(\omega_f^{-1} - \omega_i^{-1} ) t / \tau]$, $\dot\omega(t)/\omega^2(t) = (\omega_{i}^{-1} - \omega_{f}^{-1})/\tau$ is constant and the vector $\vec\phi^{\rm k}(t)$ can be expressed in closed form at any $t$. 

In the quantum isochore, the matrix $\mathcal{M}$ of Eq.~\eqref{eq:adiabaticHLDI} is given as 
\begin{align}
    \mathcal{M}^{\bf Q}_{\bf i}
    =
    \begin{pmatrix}
    -\frac{\gamma }{2} & 0  & 0 & \frac{\gamma \tilde T_{\bf i} }{2}  \\
    0 &  -\frac{\gamma }{2} & - 2\omega_{\bf i} &  0  \\
    0 & 2\omega_{\bf i}          & -\frac{\gamma }{2} & 0 \\
    0 & 0 & 0 & 0
    \end{pmatrix},
\label{eq:qiso_mat}
\end{align}
and for the classical isochoric process, the matrix is given as
\begin{align}
    \mathcal{M}^{\bf C}_{\bf i}
    =
    \begin{pmatrix}
    -\frac{\gamma }{4} & -\frac{\gamma }{4} & 0 & \frac{\gamma  T_{\bf i} }{4}\\
    -\frac{\gamma }{4} &  -\frac{\gamma }{4} & - 2\omega_{\bf i} & \frac{\gamma  T_{\bf i} }{4}  \\
    0 & 2\omega_{\bf i} & -\frac{\gamma }{4} & 0 \\
    0 & 0 & 0 & 0
    \end{pmatrix}
\label{eq:ciso_mat}
,\end{align}
where the subscript $\bf i\in\{\rm c$, $\rm h\}$ denotes either cold (c) or hot (h) isochore, and we set $k_{\rm B}=1$ for simplicity. For the case of a quantum heat bath~\eqref{eq:qiso_mat}, the Hamiltonian is decoupled from the Lagrangian and the correlation and directly approaches the equilibrium value $\tilde T_{\bf i}$. On the other hand, with a classical heat bath, the Hamiltonian, Lagrangian, and correlation remain coupled, and thus, the Hamiltonian can show oscillating behavior rather than directly approaching the equilibrium value~$T_i$.

Combining these matrices, the one-cycle propagator is written as 
\begin{align}
    {\mathcal P}^{\rm k} =&e^{\mathcal{M}^{\rm k}_{\rm c} \tau_{\rm c} }
    e^{\int^{\tau_{\rm hc }}_{0} {\rm d}t \mathcal{M}_{\bf exp}  }
    e^{\mathcal{M}^{\rm k}_{\rm h} \tau_{\rm h} }
    e^{\int^{\tau_{\rm ch }}_{0} {\rm d}t \mathcal{M}_{\bf com}  }
\label{eq:prop}
\end{align}
with the superscript $\rm k\in \{\bf Q, \bf C\}$. Here $\tau_{\rm c}$, $\tau_{\rm ch}$, $\tau_{\rm h}$, and $\tau_{\rm hc}$ stand for the time of the cold isochore, the time of the adiabatic compression process, the time of the hot isochore, and the time of the adiabatic expansion process.
The condition for the cyclic steady state (limit cycle, ss) is that state $\vec\phi_{\rm ss}(t_0)$ has to return to its initial condition after one cycle of $\tau_{\rm cyc}\equiv \tau_{\rm ch}+\tau_{\rm h}+\tau_{\rm hc}+\tau_{\rm c}$, {\em i.e.} $\vec \phi_{\rm ss} (t_0) = \mathcal P^{\rm  k} \vec \phi_{\rm ss} (t_0).$ 
From this condition, the cyclic steady state is derived as
\begin{align}
    \vec \phi_{\rm ss}^{\rm k}(t_0) = (
    c_1^{\rm k}, c_2^{\rm k}, c_3^{\rm k}, 1)
    \label{eq:firstmomentdyna}
\end{align}
where 
\begin{align}
    \vec c^{\rm k} = (\mathcal{I} - \mathcal{G}^{\rm k} )^{-1} \cdot \vec b^{\rm k},
\end{align}
\begin{align}
    \mathcal{G}^{\rm k}
    =
    \begin{pmatrix}
    P_{1,1}^{\rm k} & P_{1,2}^{\rm k} & P_{1,3}^{\rm k}  \\
    P_{2,1}^{\rm k} & P_{2,2}^{\rm k} & P_{2,3}^{\rm k}   \\
    P_{3,1}^{\rm k} & P_{3,2}^{\rm k} & P_{3,3}^{\rm k} 
    \end{pmatrix}
    \text{ and } 
    \vec{b}^{\rm k}
    =
    \begin{pmatrix}
    P_{1,4}^{\rm k} \\
    P_{2,4}^{\rm k}  \\
    P_{3,4}^{\rm k}
    \end{pmatrix}
.\end{align}

The Otto cycle can be seen as a periodic system with a driving force. So if the system lacks dissipation in the isochoric processes, then the system is divergent, which means that the largest eigenvalue of ${\mathcal P}^{\rm k}$ is larger than $1$~\cite{Insinga2018quantum,Lee2020Finite}. 
Divergent behavior can be seen near the resonance conditions~\cite{Lee2020Finite}
\begin{align}
    \begin{split}
    n \pi =& \Delta \theta_{\rm cyc} =\int^{\tau_{\rm cyc }}_{0} {\rm d}t \, \omega(t) \\
    =& \omega_{\rm c}\tau_{\rm c} + \omega_{\rm h}\tau_{\rm h} + \frac{\omega_{\rm c}\omega_{\rm h}}{\omega_{\rm h} - \omega_{\rm c}}\ln{\left( \omega_{\rm h}/\omega_{\rm c} \right)}(\tau_{\rm ch} + \tau_{\rm hc} )
    \end{split}
    \label{eq:resonance}
\end{align} where $n$ is an integer.
Because of the left-right symmetry in the Otto cycle, the left-hand side of Eq.~\eqref{eq:resonance} is given as $n\pi$, not $2 n \pi$. 
With the cyclic steady state $\vec \phi_{\rm ss}(t)$ that corresponds to eigenvalue $1$, the thermodynamic quantities of the Otto cycle such as efficiency ($\eta$), power ($P$), and entropy production ($\Sigma$) can be calculated, and the thermodynamic quantities show interesting phenomena near the resonance conditions~\cite{Kosloff2014-Review, Lee2020Finite}. 

The two-time correlation functions can be calculated in a similar way. The quantum regression theorem states that if the one-time correlation functions are governed by a linear equation, then the two-time correlation functions are also governed by the same linear equation~\cite{Breuer2006TheTheory}. Thus, the governing equation of the two-time correlation functions is written as 
\begin{align}
    \frac{d}{d t} \vec{C}^{\rm k} (t,t') = \mathcal{M}_{\rm j}^{\rm k} \vec{C}^{\rm k}(t,t')
\end{align}
where $\rm k\in\{\bf Q, C\}$, $\rm j\in\{{\bf com, exp}, c, h\}$, and
\begin{align}
\begin{split}
    \vec C^{\rm Q}(t,t') =& 
    \langle(  \hat{H}(t) \hat{H}(t') ,
 \hat{L}(t) \hat{H}(t') ,
 \hat{D}(t) \hat{H}(t'), \hat{H}(t') )\rangle_{\rm s},\\
 \vec C^{\rm C}(t,t') =& 
    \langle(  H(t) H(t') ,
 L(t) H(t') ,
 D(t) H(t'), H(t') )\rangle
 .\end{split}
\nonumber
\end{align} 
If we know the initial condition $\vec C(t', t') $, then $\vec C(t,t') $ is calculated by applying the same propagators of the one-time correlation functions.
Due to the Gaussian property of the cyclic steady state, the second moments of bases $\hat H$, $\hat L$, and $\hat D$ can be written in terms of the first moments as follows:
\begin{align} \label{eq:Hsq}
    \langle \hat H^2(t) \rangle = 2 \langle \hat H(t) &\rangle^2 + \langle \hat L(t) \rangle^2 + \langle \hat D(t)\rangle^2 - \frac{\hbar^2\omega^2(t)}{4} \\
    \langle \hat L(t) \hat H(t) \rangle_{\rm s} =& 
    3 \langle \hat H(t) \rangle \langle \hat L(t) \rangle  \\
    \langle \hat H(t) \hat D(t) \rangle_{\rm s}=& 3 \langle \hat D(t) \rangle \langle \hat H(t) \rangle 
.\end{align}
The last term of Eq.~\eqref{eq:Hsq} originates from the quantum uncertainty relation $[\hat p, \hat x] = \hbar/i$, and the negative term ensures that energy fluctuations become zero when all states are in the ground state of energy $\hbar\omega/2$. Classical relations are obtained by setting the operators as numbers in the classical limit $\hbar \rightarrow 0$. 

\section{Result}
\label{result}


\subsection{Quasistatic result}

In the quasistatic limit, the working fluid is in Boltzmann form at any time, which means no coherence and no dependency on the structure of the governing equation. The difference between quantum and classical Otto cycles in the quasistatic limit is mainly attributed to Bose--Einstein statistics and the quantum uncertainty relation, $[\hat p, \hat x]=\hbar/i$.
From the joint probability, Eq.~\eqref{eq:joint}, we calculate the mean values of the thermodynamic quantities as follows:
\begin{align}
    \langle \hat W\rangle 
    =&  (\hbar\omega_{\rm h} - \hbar\omega_{\rm c}  ) \langle \hat n_h - \hat n_c \rangle   \label{eq:W}\\
    \langle \hat Q_{h} \rangle =& \hbar\omega_{\rm h}  \langle \hat n_{h} - \hat n_{c}\rangle  \label{eq:Qh}\\
    \langle \hat Q_{c} \rangle =&  \hbar\omega_{\rm c} \langle \hat n_{h} - \hat n_{c}\rangle \label{eq:Qc}\\
    \langle \hat \Sigma \rangle     =&  (\beta_c \hbar\omega_c -  \beta_h \hbar\omega_h )\langle \hat n_h - \hat n_c \rangle  \label{eq:ent}
\end{align}
where $ \langle \hat n_{\rm h}   -  \hat n_{\rm c} \rangle  = \frac{1}{2}(\coth{(\frac{\beta_{\rm h}\hbar\omega_{\rm h}}{2})} - \coth{(\frac{\beta_{\rm c} \hbar \omega_{\rm c}}{2}) })$.
The above quantities depend linearly on the quanta difference $ \langle \hat n_{\rm h}  -  \hat n_{\rm c} \rangle$, and substituting $\hbar \langle \hat n_{\rm h}  - \hat n_{\rm c} \rangle$ in Eqs.~\eqref{eq:W}--\eqref{eq:ent} with $\frac{1}{\omega_{\rm h} \beta_{\rm h} } - \frac{1}{\omega_{\rm c} \beta_{\rm c} }$ yields the classical results. 
Due to the conservation of the number of quanta or the action in the adiabatic process, the efficiency ($\eta$) only depends on the frequencies ($\omega_{\rm h}$, $\omega_{\rm c}$), regardless of either the quantum or classical Otto cycle. 

The efficiency in the quasistatic limit is called the Otto efficiency and is written as
\begin{align}
    \eta_{_{\rm O}} = 1 - \omega_{\rm c}/\omega_{\rm h}.
    \label{eq:Ottoeffi}
\end{align}
The Otto efficiency, Eq.~\eqref{eq:Ottoeffi} is less than the Carnot efficiency ($\eta_{_{\rm C}}=1-T_{\rm c}/T_{\rm h}$).
If $\omega_{\rm c}/\omega_{\rm h} < T_{\rm c}/T_{\rm h} $, then the Otto cycle works as a refrigerator with a cooling coefficient of performance 
\begin{align}
    \delta_{_{\rm O}} = \frac{\omega_{\rm c}}{\omega_{\rm h}-\omega_{\rm c}}
.\end{align}
In the quasistatic limit, the power of the Otto cycle is zero, because the total cycle time $\tau_{\rm cyc}$ is infinite.

With one-time and two-time correlation functions, the fluctuations of the thermodynamic quantities are calculated as: 
\begin{align}
    \langle \Delta \hat W^2 \rangle =& \hbar^2(\omega_h - \omega_c)^2
    (\langle \Delta \hat n_{\rm c}^2\rangle  +\langle\Delta \hat n_{\rm h}^2\rangle) \\
    \langle \Delta \hat Q_h^2 \rangle =& \hbar^2\omega_{h}^2 
    (\langle \Delta \hat n_{\rm c}^2\rangle + \langle \Delta \hat n_{\rm h}^2\rangle) \\
    \langle \Delta \hat Q_c^2 \rangle =& \hbar^2\omega_{c}^2 
    (\langle \Delta \hat n_{\rm c}^2\rangle + \langle \Delta \hat n_{\rm h}^2\rangle) \\
    \langle \Delta \hat \Sigma^2 \rangle 
    =& \hbar^2(\beta_h  \omega_h - \beta_c  \omega_c )^2 (\langle \Delta \hat n^{2}_c \rangle + \langle \Delta \hat n^{2}_h \rangle ) 
\end{align}
where $\langle \Delta \hat n_{\rm c}^2 +\Delta \hat n_{\rm h}^2\rangle = [\coth^2{(\frac{\beta_{\rm h}\hbar\omega_{\rm h}}{2})} +\coth^2{(\frac{\beta_{\rm c}\hbar\omega_{\rm c}}{2})}- 2]/4  $.
By substituting $\hbar^2 \langle \Delta \hat n_{\rm c}^2 +\Delta \hat n_{\rm h}^2\rangle$ with $ \frac{1}{\beta_{\rm h}^2 \omega_{\rm h}^2 } + \frac{1}{\beta_{\rm c}^2 \omega_{\rm c}^2 }$, we get the fluctuations of the classical Otto cycle.
Based on the fact that $\hbar^2\langle \Delta \hat n_{\rm c}^2 +\Delta \hat n_{\rm h}^2\rangle$ is smaller than $ \frac{1}{\beta_{\rm h}^2 \omega_{\rm h}^2 } + \frac{1}{\beta_{\rm c}^2 \omega_{\rm c}^2 }$ and $\hbar |\langle \hat n_{\rm h}  - \hat n_{\rm c} \rangle| $ is smaller than $ |\frac{1}{\omega_{\rm h} \beta_{\rm h} } - \frac{1}{\omega_{\rm c} \beta_{\rm c} }|$, we derive the following two relations between classical and quantum thermodynamic variables
\begin{align}
    \langle \Delta \hat A^2 \rangle  \leq&  \langle \Delta A^2 \rangle \label{eq:fluc_inequality}\\
    |\langle \hat A \rangle | \leq& | \langle A \rangle  |
    \label{eq:current_inequality}
\end{align}
where $\hat A \in \{ \hat W, \hat Q_{\rm h}, \hat Q_{\rm c}, \hat \Sigma \}$. 

\begin{figure}
 \includegraphics*[width=\columnwidth]{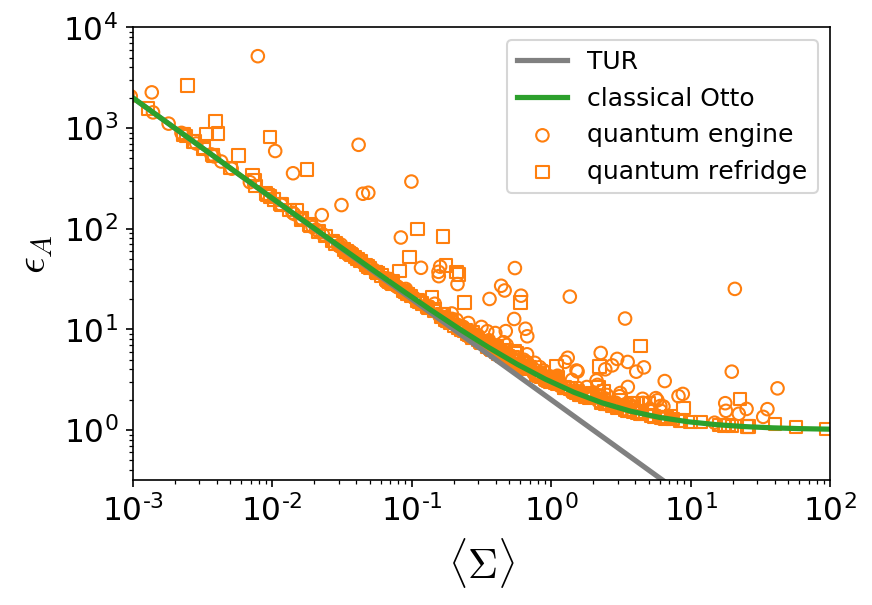}
\caption{In the quasistatic limit ($\tau_{\rm cyc}\rightarrow \infty$), the scatter plots of relative errors ($\epsilon_{\hat A}$) are shown as the function of the entropy production ($\langle \Sigma\rangle$), For the relative errors of the classical Otto cycle, $\epsilon_{A}=1+2/\langle \Sigma\rangle$ from Eq.~\eqref{eq:inequal}, which are shown as the green solid line, while those of the quantum Otto cycles are plotted with orange symbols. The bound of the conventional TUR, $\epsilon_A = 2/\langle \Sigma\rangle$, is drawn as the black solid line. }
\label{fig:quasiTUR}
\end{figure}

Regarding the relative error, we conjecture $\frac{\langle \Delta \hat n_{\rm c}^2 + \Delta \hat n_{\rm h}^2 \rangle }{\langle \hat n_{\rm h} - \hat n_{\rm c} \rangle^2} \geq \frac{\beta_{\rm h}^2\omega_{\rm h}^2 + \beta_{\rm c}^2\omega_{\rm c}^2 }{(\beta_{\rm h}\omega_{\rm h}-\beta_{\rm c} \omega_{\rm c})^2}$, which implies that the classical Otto cycle is more reliable than the quantum Otto cycle by
\begin{align}
    \frac{\langle \Delta \hat A^2 \rangle}{\langle \hat A \rangle^2}  \geq  \frac{\langle \Delta A^2 \rangle}{\langle A \rangle^2}  
.\end{align}
The related mathematical proofs and evidence for the inequalities are provided in Appendix~\ref{appendix:A}.
Even though the quantum uncertainty relation results in the quantum fluctuations $\langle \Delta\hat A^2 \rangle $ being lower than the classical ones $\langle \Delta A^2 \rangle$, due to the smaller mean value, the quantum relative error is larger than the classical one. These results originate from the Bose--Einstein statistics and uncertainty relation. Regarding the productivity, we provide an intuitive explanation as follows: Due to Bose--Einstein statistics, the slope of $\hbar\langle\hat{n}\rangle$ is always smaller than $\hbar\langle \hat{n}\rangle$ in the classical limit, $\hbar \to 0$. Because the work of the quantum Otto cycle is proportional to the difference of the quanta at two isochores, $\hbar(n_{\rm h} -n_{\rm c})$, and the classical work is yielded by taking the classical limit, the classical work is always larger than the quantum one. 

For a test of the conventional TUR, the relative error, $\epsilon_A$, is written as 
\begin{align}
    \epsilon_{\hat A}
    = 1 + U \frac{2 }{\langle \hat \Sigma \rangle }
    \label{eq:quasiQrel}
\end{align}
where
\begin{align}
    U \equiv \frac{\beta_{\rm c}\hbar \omega_{\rm c } - \beta_{\rm h}\hbar\omega_{\rm h} }{2} 
    \coth{\left(\frac{\beta_{\rm c}\hbar \omega_{\rm c } - \beta_{\rm h}\hbar\omega_{\rm h} }{2}\right)}
.\end{align}
We plot the relative errors, Eq.~\eqref{eq:quasiQrel}, in Fig.~\ref{fig:quasiTUR}. Provided that $U\geq 1$, we can write an inequality about the relative error,
\begin{align}
    \epsilon_{\hat A} \geq 1+ \frac{2}{\langle \hat \Sigma \rangle}
    \label{eq:inequal}
.\end{align}
Most recently, the same result of this inequality [Eq.~\eqref{eq:inequal}] has been reported in a two-mode bosonic Otto engine~\cite{sacchi2020thermodynamic} that uses two working fluids.  In the classical limit ($\hbar \rightarrow 0$), Eq.~\eqref{eq:quasiQrel} becomes $\epsilon_{A}= 1 +\frac{2}{\langle \Sigma  \rangle }$. As a result, the inequality of Eq.~\eqref{eq:inequal} is tight for the classical Otto cycle, and in the conventional TUR, $\epsilon_{A} \geq \frac{2}{\langle \Sigma \rangle}$, is valid (see Fig.~\ref{fig:quasiTUR}).

\subsection{Finite-time result}

In the finite-time mode, the results are quite different from the quasistatic results.
\begin{figure*}[]
\includegraphics*[width=\textwidth]{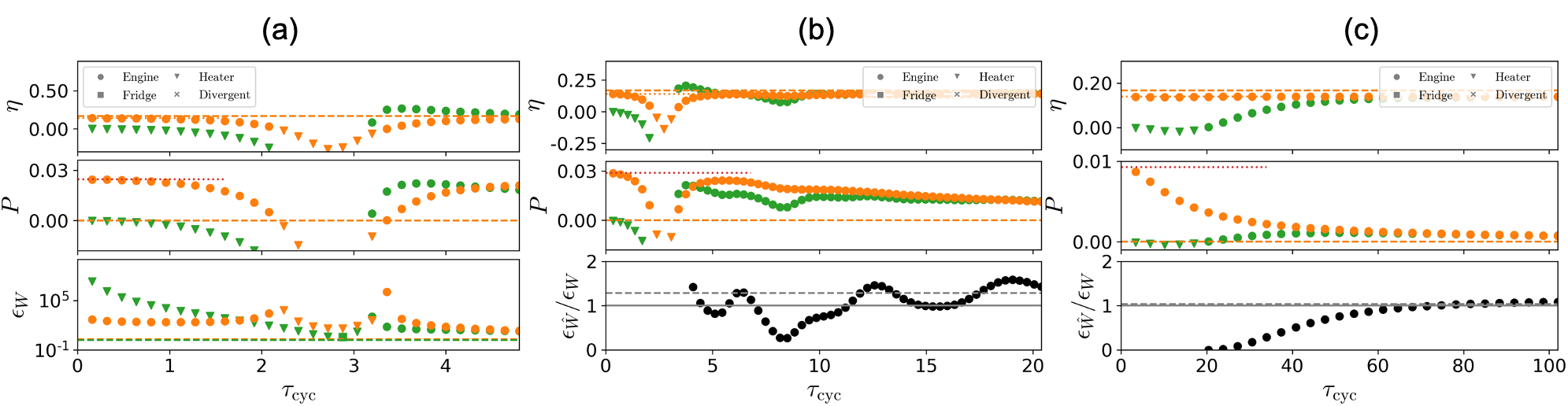}
\caption{For both quantum and classical Otto cycles in the finite-time mode, we plot from top to bottom the efficiency ($\eta$), the power ($P$), and the relative errors of work ($\epsilon_{W}, \epsilon_{\hat{W}}$) in (a) [the rescaled relative errors of work (the quantum relative errors $\epsilon_{\hat{W}}$ divided by the classical one $\epsilon_{W}$) are plotted in (b) and (c) only for the engines] as a function of $\tau_{\rm cyc}$ with $\gamma=m=1$, where orange symbols correspond to the quantum case, green symbols to the classical case, and black symbols to the ratio of the quantum case to the classical case, respectively: For (a) and (b), $\omega_{\rm h } = 1.2, \omega_{\rm c}=1.0, T_{\rm h}=2$, and $T_{\rm c}=0.3$, and for (c), $\omega_{\rm h}=0.12, \omega_{\rm c}=0.1, T_{\rm h} =0.4$, and $T_{\rm c}=0.1$. For all three cases, the ratio of the adiabatic process time ($\tau_{\rm ch}=\tau_{\rm hc}$) to the isochoric time ($\tau_{\rm c}=\tau_{\rm h}$) is $\tau_{\rm hc}/\tau_{\rm c}=0.2$. 
In particular for (a), we check the short-time regime and also show all possible types (different symbols) of the Otto cycle in the bottom plot of the relative work errors. Here dashed lines correspond to the asymptotic limiting values, and dotted lines correspond to the short-time limiting values of the quantum Otto cycle. For the rescaled relative errors, we observe that they can be smaller than 1 in the short-time regime, which implies that the quantum engine is more precise than the classical engine.}
\label{fig:effpower}
\end{figure*}
In Fig.~\ref{fig:effpower}, we plot the finite-time efficiency, power, and relative error of the Otto cycles. The finite-time Otto cycle can show divergent or oscillatory behaviors near the resonance conditions~\eqref{eq:resonance}. For the quantum beat bath, coherence ($\hat L$, $\hat D$) and energy ($\hat H$) are decoupled, and thus the bath merely dissipates the working fluid energy, which previously increased with increasing coherence in the adiabatic process. The efficiency of the Otto cycle with the quantum bath, therefore, does not exceed the quasistatic Otto efficiency. This phenomenon is called quantum friction~\cite{Rezek2006Irreversible, Kosloff2017-QuantumOtto}. However, the Otto engine with a classical or Agarwal bath shows higher efficiency than the quasistatic efficiency~\cite{Lee2020Finite} in the vicinity of resonance conditions. Similar efficiency enhancement was observed in the finite-time Stirling cycle~\cite{hamedani2020finite}. 

We extract energy from the harmonic potential that only depends on the position variable, with which the quantum thermostat is in direct contact. Because of this reason, the quantum Otto cycle can work as an engine even in the short-time limit $\tau_{\rm cyc} \rightarrow 0 $. Conversely, the classical thermostat only contacts to the momentum, and consequently the classical Otto cycle does not produce work in this short time limit. 

The relative error (relative fluctuation) is another important measure for the thermodynamic quantities of the thermal devices. In the bottom panel of Fig.~\ref{fig:effpower}, we plot (a) the measure of relative errors, $\epsilon_{\hat{W}} \equiv \langle \Delta \hat W^2 \rangle /\langle \hat W\rangle^2$ for the quantum case and $\epsilon_{W}\equiv \langle \Delta W^2\rangle/\langle W\rangle$ for the classical case, and (b) and (c) their ratio (the quantum relative errors to the classical relative error) when the Otto cycle performs as a heat engine. Thus, on  account of quantumness, the quantum Otto cycle can produce work and is more reliable than its classical counterpart. 

\begin{figure}
 \includegraphics*[width=\columnwidth]{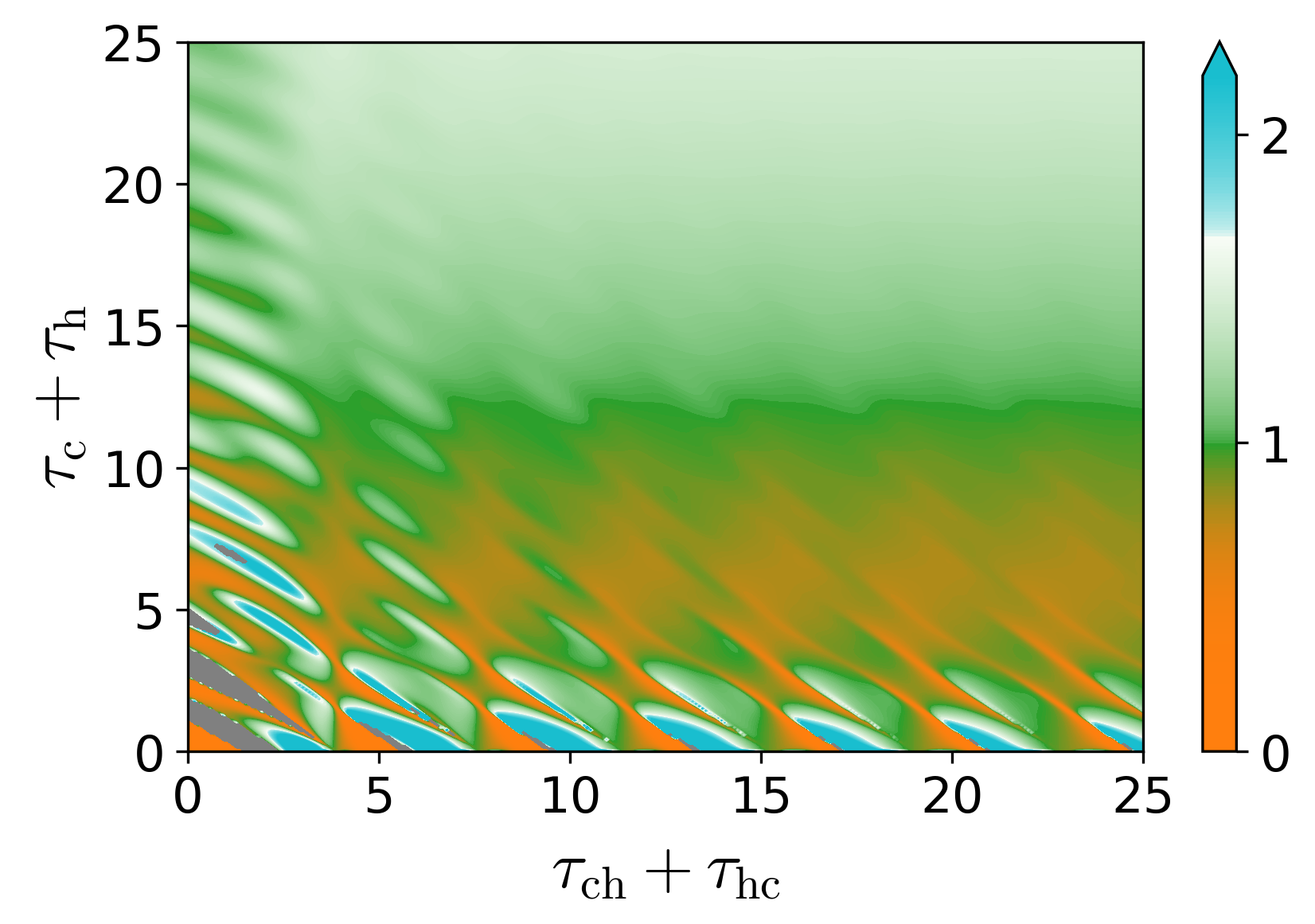}
\caption{The contour plot of the ratio of quantum to classical relative errors of work, $\epsilon_{_{\hat W}}/\epsilon_{_{W}}$, is presented with the time of adiabatic processes, $\tau_{\rm ch}+\tau_{\rm hc}$, and the time of isochoric processes, $\tau_{\rm h}+\tau_{\rm c}$ , where the orange-colored region corresponds to the case when the quantum relative error is smaller than its classical counterpart and the green-colored region to the opposite case. The colored region continuously changes to white as the relative error approaches the quasistatic value. If the ratio is above the quasistatic value, then the data are colored by light blue. When either the quantum Otto cycle or the classical one shows divergent behaviors, we color the region as gray. Here we use the following parameters: $m=1$, $\gamma = 1$, $\omega_{\rm h} = 3$, $\omega_{\rm c} = 1 $, $T_{\rm h}=1$, and $T_{\rm c}=0.5$. We also set $\tau_{\rm ch}=\tau_{\rm hc}$ and $\tau_{\rm c}=\tau_{\rm h}$. 
}
\label{fig:contour}
\end{figure}

\begin{figure}
 \includegraphics*[width=0.85\columnwidth]{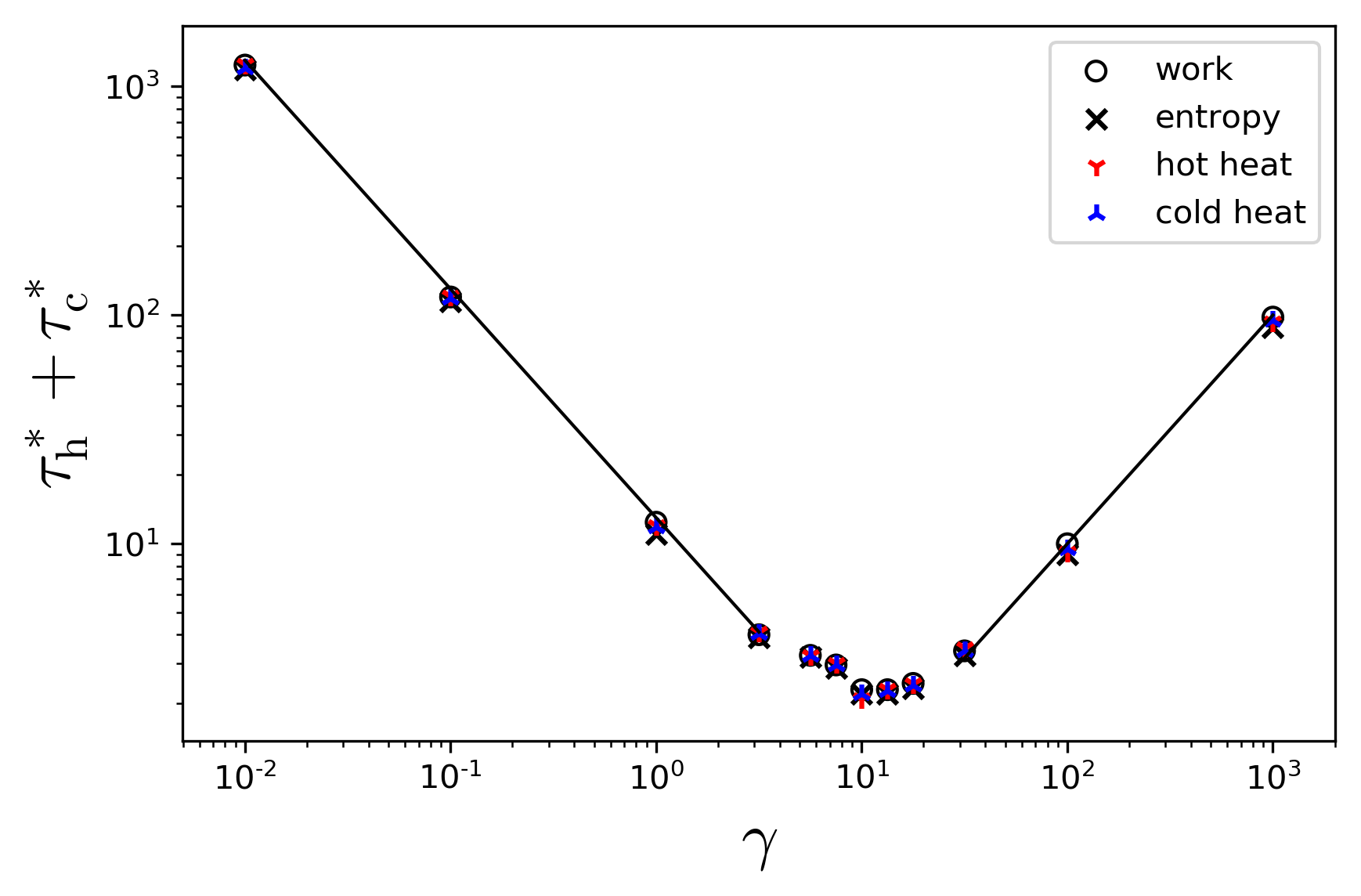}
\caption{ The $\gamma$ dependency of the value of the total isochoric time, $\tau^*_{\rm h} + \tau^*_{\rm c}$, above which the ratio of the quantum relative error to the classical one is smaller than 1, is plotted. 
{\em i.e. }, $\epsilon_{\hat A}/\epsilon_{A}=1$. Here ${\hat A}\in\{{\hat W}, {\hat \Sigma}, {\hat Q}_{\rm h}, {\hat Q}_{\rm c}\}$ and $A\in\{W, \Sigma, Q_{\rm h}, Q_{\rm c}\}$. We use the same parameters as those in Fig.~\ref{fig:contour}. The error bar of numerical data is roughly the same as the size of the symbols, the slope of the solid line on the left is -1 ($\propto \gamma^{-1}$), and the slope of the solid line on the right is 1 ($\propto \gamma$).}
\label{fig:scaling}
\end{figure}

\begin{figure*}
 \includegraphics*[width=\textwidth]{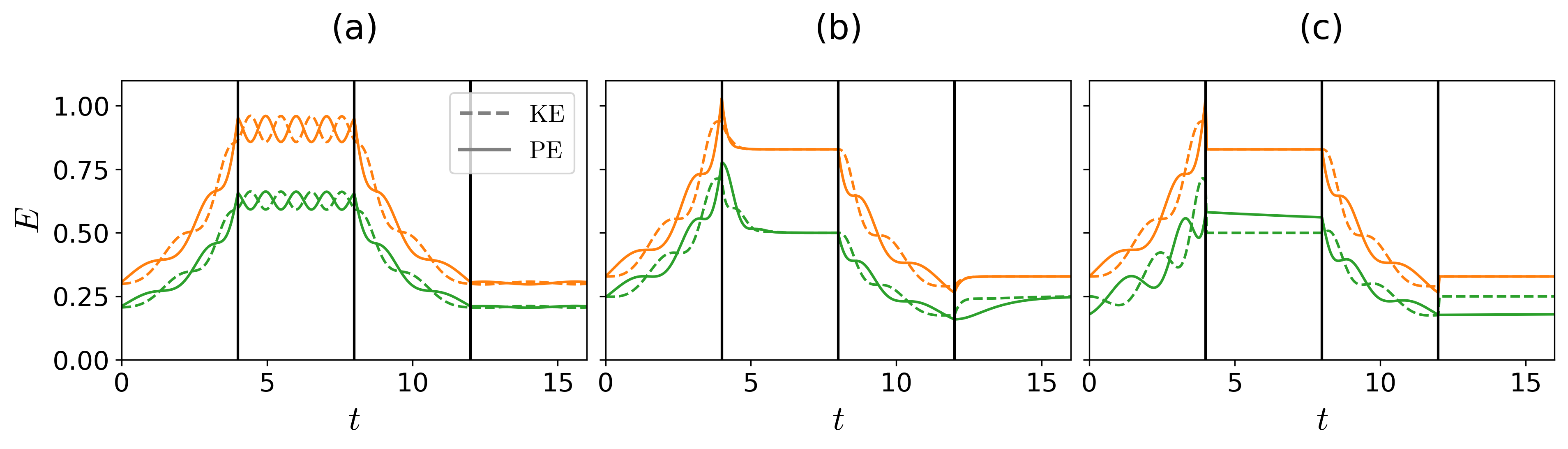}
\caption{ Trajectories of the kinetic energy (KE) and the potential energy (PE) during one period are plotted for three different values of heat conductance: (a) $\gamma=0.01$, (b) $\gamma = 10 $, and (c) $\gamma = 1000$. The vertical black lines divide the compression process~$0\leq t< 4$, the hot isochoric process~$4\leq t< 8$, the compression process~$8\leq t < 12$, and the cold isochoric process~$12\leq t< 16 $. In the high heat conductance regime, we observe that the classical Otto cycle (colored in green) is in a nonequilibrium state, but the quantum Otto cycle (colored in orange) is in an equilibrium state, see (c) at the end of thehot and cold isochores.}
\label{fig:KEPE}
\end{figure*}

Then, we arrive at the following question: ``When does the quantum Otto cycle start to become more reliable (with less relative errors of work) than the classical Otto cycle?". Figure~\ref{fig:contour} shows a contour plot of the ratio of quantum to classical relative errors of work, $\epsilon_{_{\hat W}}/\epsilon_{_{W}}$, as a function of the summation of two adiabatic process times and the summation of two isochoric process times only when neither Otto cycle diverges. 

In Fig.~\ref{fig:contour}, the orange regions represent that the quantum Otto cycle is more reliable than the classical cycle, while the green regions represent that the classical Otto cycle is more reliable than the quantum one. If the ratio is over the quasistatic value, then we color it with light blue. The oval pattern originates from resonance phenomena, Eq.~\eqref{eq:resonance} and this pattern is determined by the frequency of the harmonic potential~\cite{Insinga2016Thermodynamical}. 

From Fig.~\ref{fig:contour}, we observe that the orange-colored region is located only below the certain value of $\tau_{\rm c}+\tau_{\rm h}$ (the total isochoric time), $\tau_{\rm c}^*+\tau_{\rm h}^{*}$, which depends on the value of the heat conductance $\gamma$ and is almost independent of $\tau_{\rm ch}+\tau_{\rm hc}$ (the total adiabatic time). To discuss the region where the quantum relative error becomes smaller than the classical one, we measure the $\gamma$ dependency of $\tau_{\rm c}^*+\tau_{\rm h}^*$. In Fig.~\ref{fig:scaling}, without loss of generality, we fix the total adiabatic time and take $\tau_{\rm c}^*+\tau_{\rm h}^*$ as the value of the total isochoric time, above which no orange color is observed~\footnote{More specifically, the ratio of the quantum relative error to the classical one is smaller than 1}.

In the finite-time mode, the difference between the quantum and classical Otto cycles originates from how coherence~($\hat L$, $\hat D$) is dealt. With a long isochoric time, coherence disappears and we expect that the classical Otto cycle is more reliable than the quantum one as in the quasistatic case. The criterion to determine what constitutes a long isochoric time is found by the inverse of heat conductance~$4/\gamma$. Thus, we infer that $\tau_{\rm h}^{*}+\tau^{*}_{\rm c}$ is inversely proportional to $\gamma$.
We find that $\tau^{*}_{\rm c} + \tau^{*}_{\rm h}$ is inversely proportional to $\gamma$ when $\gamma$ is less than $1$ in Fig.~\ref{fig:scaling}. 

Interestingly, in the high $\gamma$ regime~($\gamma >10$), $\tau^{*}_{\rm c} + \tau^{*}_{\rm h}$ starts to increase, which means that the region where the quantum Otto cycle is more reliable than the classical one is expanded. 
Because heat conductance is the strength between the heat bath and the system, it seems counter-intuitive that the result does not approach the quasistatic result as the heat conductance~$\gamma$ increases. We find the reason from the governing equation for the kinetic energy and the potential energy. The governing equations for the classical Otto cycle are written as 
\begin{align}
    \frac{\rm d}{{\rm d}t} \langle \frac{ p ^2 }{2m}\rangle &= - \frac{\gamma}{2} \langle   \frac{ p ^2 }{2m}\rangle - \omega_{\bf i} D + \frac{\gamma T_{\bf i}}{4}, \label{eq:KE}\\
    \frac{\rm d}{{\rm d}t} \langle \frac{m \omega^2_{\bf i}  x ^2 }{2}\rangle &=  \omega_{i} D,\label{eq:PE}\\
    \frac{\rm d}{{\rm d}t} D &= 2\omega_{\bf i } \langle  \frac{p^2}{2m}\rangle -2\omega_{\bf i } \langle \frac{m\omega^2_{\bf i}x^2}{2} \rangle -\frac{\gamma}{4} D.
  \label{eq:PEKE_C}
\end{align}
While the kinetic energy is directly equilibrated by the thermostat, Eq.~\eqref{eq:KE}, the potential energy is indirectly equilibrated via the correlation ($D$) between position and momentum, Eq.~\eqref{eq:PE}. In the high~$\gamma$ region, Eqs.~\eqref{eq:KE}, \eqref{eq:PE}, and \eqref{eq:PEKE_C} are written as 
\begin{align}
    \frac{\rm d}{{\rm d}t} \langle \frac{ p ^2 }{2m}\rangle &\simeq - \frac{\gamma}{2} \langle   \frac{ p ^2 }{2m}\rangle + \frac{\gamma T_{\bf i}}{4}, \\
    \frac{\rm d}{{\rm d}t} \langle \frac{m \omega^2_{\bf i}  x ^2 }{2}\rangle &\simeq  \omega_{i} D\\
    \frac{\rm d}{{\rm d}t} D &\simeq -\frac{\gamma}{4} D.
  \label{eq:PEKE_C_high_gamma}
\end{align}
In this high~$\gamma$ limit, the kinetic energy and the correlation ($D$) approach the corresponding equilibrium values, $T_{\bf i}/2$ and $0$, respectively. When $D$ becomes $0$, the potential energy cannot reach the equilibrium value and thus it has a nonequilibrium value. On the other hand, the quantum Otto cycle can approach the equilibrium value because there exists the additional thermostat, $- \frac{\gamma}{2} \langle \frac{m \omega^2_{\bf i} \hat x ^2 }{2}\rangle  + \frac{\gamma \tilde T_{\bf i}}{4}$ for the potential energy. 

In Fig.~\ref{fig:KEPE}, we plot the trajectories of the kinetic energy and the potential energy with~$\gamma=0.01$, $10$ and $1000$ from (a) to (c). At the end of the isochoric processes~$t =8$ and $16$, there are clear differences among three panels in the figure. It can be seen that with $ \gamma = 10$, both quantum and classical Otto cycles are equilibrated at the end points. With the small heat conductance~$\gamma = 0.01$ and the short isochore time~$\tau_{\rm c} = 4$, neither cycle can be equilibrated as shown in Fig.~\ref{fig:KEPE} (c). However, with the large heat conductance~$\gamma =1000$, the classical working fluid is in a nonequilibrium state as previously explained. So the kinetic energy and the potential energy of the classical Otto cycle do not satisfy the equipartition theorem, but those of the quantum Otto cycle do. This phenomenon occurs when the heat conductance~$\gamma$ is larger than the harmonic frequency~$\omega$, which affects the equilibration speed of the potential energy of the classical cycle. This is the reason why we find that $\tau^*_{\rm h}+\tau^*_{\rm c}$ increases in Fig.~\ref{fig:scaling} when $\gamma > 10$ and the lowest value of $\tau^*_{\rm h}+\tau^*_{\rm c}$ moves to the left as the harmonic frequency decreases. 

\begin{figure}
 \includegraphics*[width=\columnwidth]{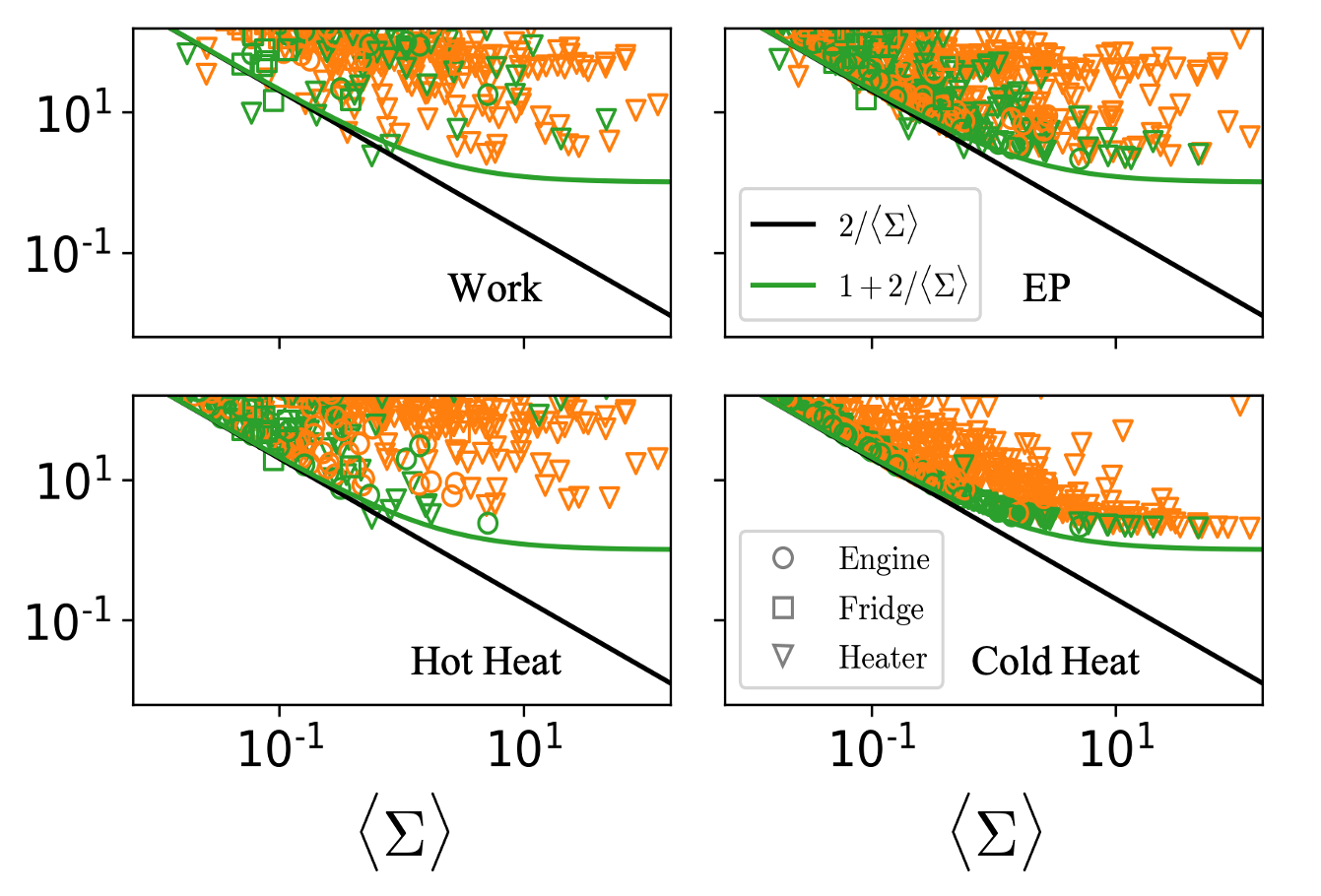}
\caption{ In the finite-time mode, the scatter plots of relative output errors are obtained with 500 combinations of frequencies and temperature bounded by 1 for $\gamma = 1/4$ and $m = 1$. In the upper two panels, the relative errors of work and entropy production (EP), $\epsilon_{{\hat A\in\{\hat W, \hat \Sigma\}}}$ and $\epsilon_{A\in\{W, \Sigma\}}$, are plotted as the function of the entropy production $\langle \Sigma\rangle$, from left and right, respectively. In the lower two panels, the relative errors of hot heat and cold heat, 
$\epsilon_{\hat A\in\{\hat Q_{\rm h}, \hat Q_{\rm c}\}}$ and $\epsilon_{A\in\{Q_{\rm h}, Q_{\rm c}\}}$, are plotted as a function of $\langle \Sigma\rangle$. Here $\tau_{\rm cyc}=5$ and the duration of each process is the same. The conventional TUR bound, $2/\langle\Sigma\rangle$, is denoted by the black solid line of each panel, and the quasistatic limiting value of the classical Otto cycle, $1+ 2/\langle\Sigma\rangle$, from Eq.~\eqref{eq:inequal} is denoted by the green solid line of each panel. Unlike the quasistatic limit of Fig.~\ref{fig:quasiTUR}, we observe that few data points of the finite-time Otto cycles are located below the conventional TUR bound in the region where the entropy production is small.}
\label{fig:finiteTUR}
\end{figure}

\begin{figure}
 \includegraphics*[width=\columnwidth]{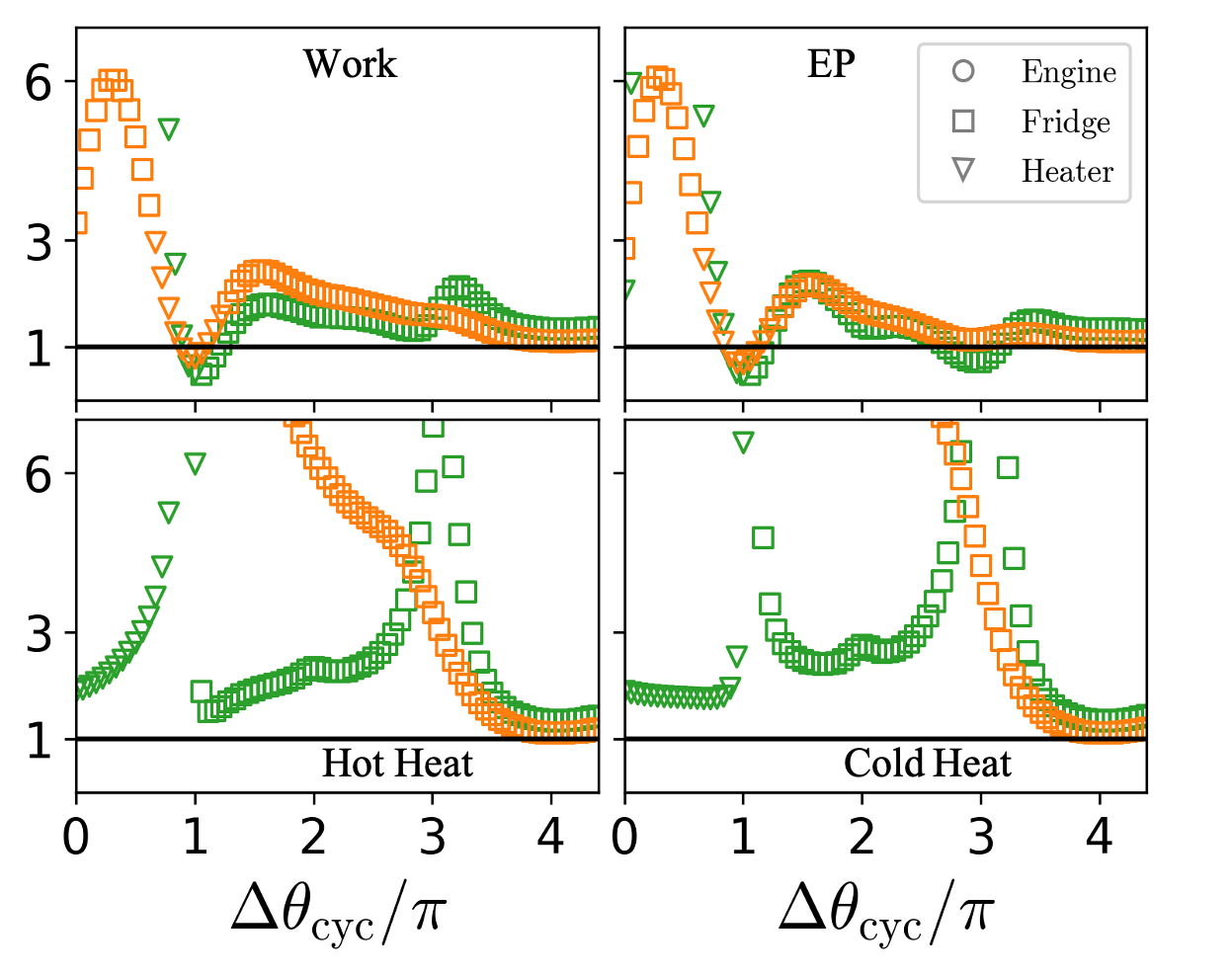}
\caption{ The uncertainty products of Eq.~\eqref{eq:def-UProduct}, namely the $\mathcal Q$ factors, $\mathcal Q_{\hat A}$ for the quantum Otto cycle (denoted as orange symbols) and $\mathcal Q_{A}$ for the classical Otto cycle (denoted as green symbols) are plotted as a function of the rescaled phase, $\Delta \theta_{\rm cyc}/\pi$, Eq.~\eqref{eq:resonance}. Here $\mathcal Q_{\hat A\in\{\hat W, \hat \Sigma, \hat Q_{\rm h}, \hat Q_{\rm c}\}}$ and $\mathcal Q_{A\in\{W, \Sigma, Q_{\rm h}, Q_{\rm c}\}}$ for work, entropy production, hot heat and cold heat from the upper-left to the lower-right panels (clockwise), respectively. Different symbols are used for engine, fridge, and heater. Since black solid lines are when ${\mathcal Q}_{\hat A}=1={\mathcal Q}_{A}$, data points below them correspond to the violation of the conventional TUR. We set parameters as $\omega_{\rm h} = 0.750... $, $\omega_{\rm c} = 0.633...$, $T_{\rm h}=0.698...$, $T_{\rm c} = 0.622...$, $m = 1$, $\gamma = 1$, and $\tau_{\rm h}=\tau_{\rm c}=\tau_{\rm ch}=\tau_{\rm hc}$.}
\label{fig:Qfactor}
\end{figure}
Finally, we investigate the TUR of a finite-time Otto cycle. In Fig.~\ref{fig:finiteTUR}, we show the scatter plots of the relative errors, $\epsilon_{\hat A}$ and $\epsilon_{A}$, as a function of entropy production $\langle \Sigma \rangle $. The diagonal black solid line represents the conventional TUR bound, and the green solid line is the quasistatic limit of the classical Otto cycle, see Eq.~\eqref{eq:inequal}. It can be seen that the finite-time Otto cycles violate the conventional TUR in the small dissipation regime, $\langle \Sigma \rangle < 10^0$, from a few points under the black solid line. On the other hand, in the large dissipation regime, $\langle \Sigma \rangle > 10^0$, the relative errors of both quantum and classical Otto engines are still bounded by~$\epsilon_{A} = 1+2/\langle \Sigma \rangle$; see Eq.~\eqref{eq:inequal}. For the classical Otto cycle, we verify one of the violated points with Monte Carlo simulations in Appendix~\ref{appendix:B}. 

In Fig.~\ref{fig:Qfactor}, we show how the duration of $\tau_{\rm cyc}$ changes the violation of the conventional TUR for the uncertainty products of work, entropy production, hot heat, and cold heat, which are defined as
\begin{align}
    \mathcal Q_{\hat A} \equiv \frac{\epsilon_{\hat A} \langle \hat \Sigma \rangle  }{2 } \text{ and }\mathcal Q_{A} \equiv \frac{\epsilon_{ A } \langle \Sigma \rangle  }{2},
    \label{eq:def-UProduct}
\end{align} 
where $\hat A \in \{ \hat W , \hat \Sigma ,\hat Q_{\rm h},\hat Q_{\rm c} \}$ and $A\in\{W, \Sigma, Q_{\rm h}, Q_{\rm c}\}$. The $x$-axis of Fig.~\ref{fig:Qfactor} is the normalized accumulated phase difference,
$\Delta \theta_{\rm cyc} / \pi = \int^{\tau_{\rm cyc}}_{0} {\rm d}t~ \omega (t)/\pi$.
 
From Fig.~\ref{fig:Qfactor}, we observe that the uncertainty products of work and entropy production abruptly decrease at which point the normalized accumulated phase becomes odd integers, but the uncertainty products of hot and cold heats abruptly increase in the vicinity of such locations because the absolute values of $Q_{\rm h}$ and $Q_{\rm c}$ decrease toward $0$. Changes in the symbols of both hot and cold heats when $\Delta \theta_{\rm cyc} / \pi = 1$ support this explanation.


\begin{figure}
 \includegraphics*[width=\columnwidth]{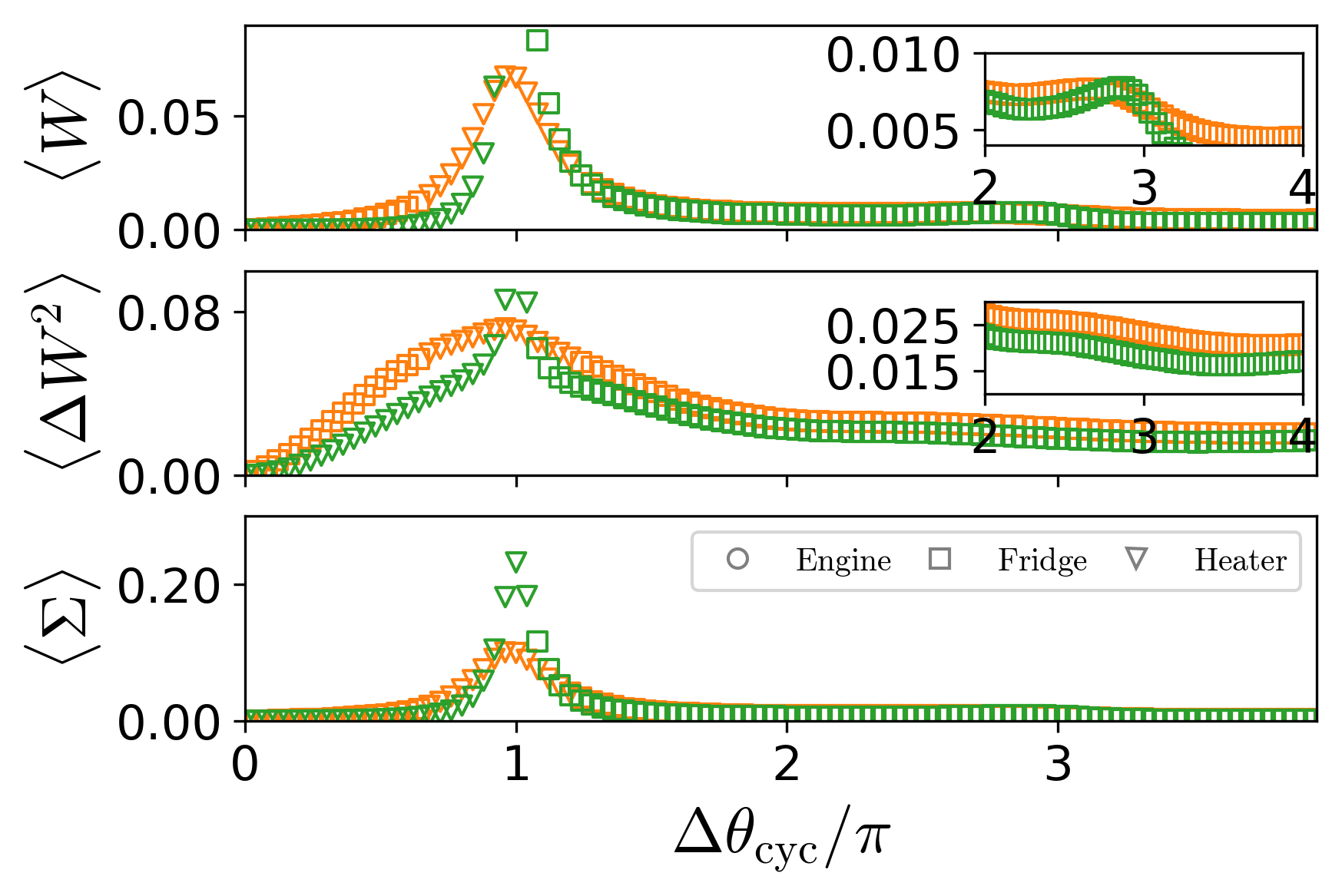}
\caption{ From top to bottom, the mean of work, the fluctuation of work, and the mean of the entropy production are plotted as a function of the normalized phase, $\Delta \theta_{\rm cyc}/\pi$. Insets are enlarged views around $\Delta \theta_{\rm cyc}/\pi = 3$. These data are the same as those used in the upper-left panel of Fig.~\ref{fig:Qfactor} for the $\mathcal Q$ factor of work.}
\label{fig:workflucent}
\end{figure}

To investigate the uncertainty product of work that shows the violation of the conventional TUR in detail, we plot the mean value of work, the fluctuation of work, and the mean value of the entropy production in Fig.~\ref{fig:workflucent}, where we find an abrupt increase of work at the resonance points. 
Since the speed of the increase of both $\langle W \rangle$ and $\langle \hat W \rangle$ is faster than that of the other quantities, we can see abrupt increments of the uncertainty product at the points. In addition, the classical Otto cycle shows more abrupt changes than the quantum Otto cycle. It seems that the additional friction caused by the positional thermostat of the quantum bath in Eq.~\eqref{eq:QCL} keeps the quantum Otto cycle from having excessive work current near the resonance conditions. Thus, we conclude that the Otto cycles produce a reliable and high work current that even violates the conventional TUR near the resonance points. 


\section{Conclusion}
\label{summary}

To reveal quantum effects on the Otto cycle, we directly compared quantum and classical cycles, where we calculated exactly the mean and fluctuations of thermodynamic quantities such as, work, hot heat, cold heat, and entropy production for the two cases. 

From these results, we found that quantumness can enhance the productivity and precision of the Otto cycle in the finite-time mode, where the working fluid has coherence. 
However, in the quasistatic limit, there is no coherence, and quantumness harms the precision and productivity of the Otto cycle. This is because of the Bose Einstein statistics and the quantum uncertainty relation.
Moreover, the relative errors of work, entropy, and hot and cold heat are all the same, and the classical Otto cycle satisfies the relation, $\epsilon_{A} = 1 +\frac{2 }{\langle \Sigma \rangle }$, which also becomes the bound of the relative errors of the quantum Otto cycle. As a result, we confirmed that neither Otto cycle violates the conventional TUR bound.

In finite-time modes, as the total cyclic time~$\tau_{\rm cyc}$ becomes shorter, the quantum working fluid becomes coherent through the adiabatic processes, and the quantumness can make thermodynamic machines more reliable. In the short time limit, $\tau_{\rm cyc}\rightarrow 0$, the positional thermostat of the quantum bath makes the Otto cycle behave as an engine. In addition, we found that the positional thermostat ensures the equilibration of the quantum Otto cycle even in the high $\gamma$ (heat conductance) limit. Furthermore, we showed that the regime where the quantum cycle is more reliable than the classical one is expanded as $\gamma$ increases with a small harmonic frequency~$\omega$. In the vicinity of the resonance points where the conventional TUR is violated, both Otto cycles show high and reliable thermodynamic currents.

Since we chose a harmonic oscillator as the working fluid, there is no difference in the quantum and classical master equations for the second moments of position and momentum in the adiabatic process. However, in general, there exist differences between the quantum and classical Otto cycles in the adiabatic process {\em e.g.} an anharmonic potential~\cite{gardiner2004quantum}. The differences might yield other quantumness effects on thermodynamic machines in the finite-time mode, which would be an intriguing area of research. 

For open quantum systems, there is no unique way in the definition of work, so that a variety of the definitions of work can be used. Among them, we employed the definition of the operational work~\cite{Breuer2006TheTheory, Kosloff2014-Review, Kosloff2017-QuantumOtto, Camati2019Coherence}, 
which can be measured for an optomechanical Otto engine using the continuous measurement~\cite{Dong2015work}. The two-point measurement is another definition of work~\cite{sacchi2020thermodynamic}, in which a couple of measurements are performed, one at the starting point and one at the end point of a process. These measurements make the coherence of the working fluid disappear. As a result, in the finite-time mode, the two-point measurement might yield a different result from ours. This may be studied in future work. 

\begin{acknowledgments} 
This research was supported by Basic Science Research Program through the National Research Foundation of Korea (NRF) (KR) [NRF-2020R1A2C1007703~(S.L., M.H.) and NRF-2017R1A2B3006930~(S.L., H.J.)], and research fund from Chosun University, 2019 (M.H. for  her sabbatical year). We thank Hyunggyu Park for helpful discussion on this research, and Hyun-Myung Chun and two Referees for kind and valuable feedbacks on the manuscript with useful references. S. L. thanks Philip Choi for helpful discussion in proving inequalities.
\end{acknowledgments}

\appendix
\renewcommand\thefigure{\thesection.\arabic{figure}}    
\setcounter{figure}{0}

\section{Inequalities for quantum and classical Otto cycles in the quasistatic limit}
\label{appendix:A}

\begin{figure}[b]
 \includegraphics*[width=\columnwidth]{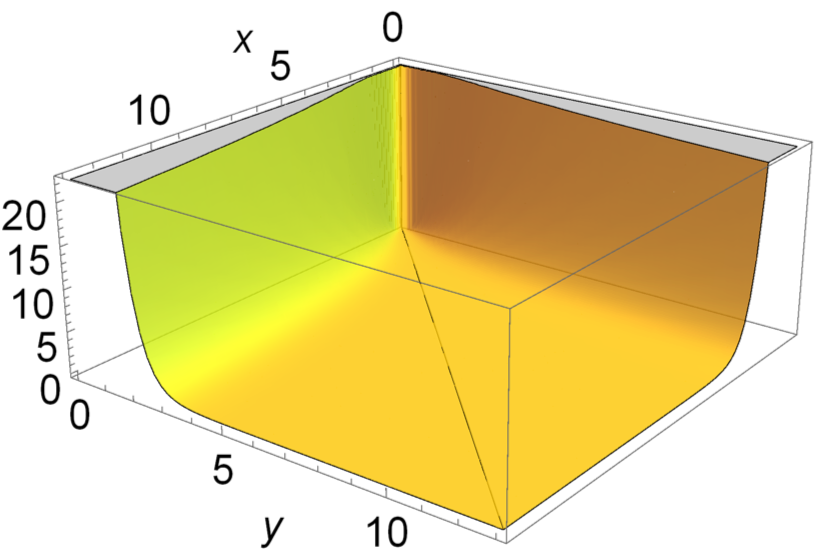}
\caption{ Contour plot of the function $l(x,y)$ which is positive in the overall region. We see that the function~$l(x,y)$ is above the plane of $z=0$.
}
\label{fig:l_poten}
\end{figure}

In this Appendix~\ref{appendix:A}, we show the proofs for three inequalities of quasistatic results:

\subsection{Proof for the inequality: $\langle \Delta \hat A^2 \rangle  \leq  \langle \Delta A^2 \rangle$}

To prove that the work fluctuations of the classical Otto cycle are larger than those of the quantum system [Eq.~\eqref{eq:fluc_inequality}], we have to prove the below inequality
\begin{align}
    \hbar^2 \langle \Delta \hat n_{\rm c}^2 +\Delta \hat n_{\rm h}^2\rangle
    \leq 
    \frac{1}{\beta_{\rm h}^2 \omega_{\rm h}^2 } + \frac{1}{\beta_{\rm c}^2 \omega_{\rm c}^2 }.
    \label{eq:appendix}
\end{align}
Equation~\eqref{eq:appendix} can be divided into two parts,
\begin{align}
    \hbar^2 \langle \Delta \hat n_{\rm c}^2 \rangle \leq \frac{1}{\beta_{\rm c}^2 \omega^2_{\rm c}}  \text{ and }
    \hbar^2 \langle \Delta \hat n_{\rm h}^2 \rangle \leq \frac{1}{\beta_{\rm h}^2 \omega^2_{\rm h}} 
.\end{align} 
We substitute $\hbar \beta_{\rm j} \omega_{\rm j} / 2$  with ${\rm j}\in\{\rm c, h\}$ as $z$, and then the inequality is rearranged into 
\begin{align}
    z - \ln{ (z + \sqrt{1+z^2} )} \geq 0
.\end{align} 
We newly define a function $g(z) \equiv z - \ln{ (z + \sqrt{1+z^2} )} $, after which the derivative of $g(z)$ is written as 
\begin{align}
    g'(z) 
    =& \frac{ z(1 - 1/\sqrt{1+z^2}) + (\sqrt{1+z^2} -1)  }{z + \sqrt{1+z^2}}\\
    \geq& 0 
.\end{align}
Because $g(0) = 0$, $g(z)$ is positive for $z\geq 0$.
Thus, the fluctuation of the quantum Otto engine is less than that of the classical one.

\subsection{Proof for the inequality: $|\langle \hat A \rangle| \leq |\langle  A \rangle|$ }

To prove Eq.~\eqref{eq:current_inequality}, the following equation has to be proved:
\begin{align}
\hbar |\langle \hat n_{\rm h}  - \hat n_{\rm c} \rangle| \leq& |\frac{1}{\omega_{\rm h} \beta_{\rm h} } - \frac{1}{\omega_{\rm c} \beta_{\rm c} }|
\\
|\coth{x} - \coth{y}| \leq& |1/x - 1/y| 
\end{align}
where $ x = \frac{\beta_{\rm h}\hbar \omega_{\rm h}}{2} $ and  $y = \frac{\beta_{\rm c}\hbar \omega_{\rm c}}{2} $. 
Without loss of generality, suppose that $x$ is greater than $y$, and then 
\begin{align}
    \coth{y}- 1/y  &\leq \coth{x} - 1/x 
.\end{align}
Because $k(x)\equiv \coth{x} - 1/x$ is an increasing function, the above equation is proved.  
\subsection{Evidence for the inequality:
$\frac{\langle \Delta \hat A^2 \rangle}{\langle \hat A \rangle^2}  \geq  \frac{\langle \Delta A^2 \rangle}{\langle A\rangle^2} $}
The inequality of the relative errors in the quasistatic limit is written as 
\begin{align}
     \frac{\langle \Delta \hat n_{\rm c}^2 + \Delta \hat n_{\rm h}^2 \rangle }{\langle \hat n_{\rm h} - \hat n_{\rm c} \rangle^2} \geq \frac{\beta_{\rm h}^2\omega_{\rm h}^2 + \beta_{\rm c}^2\omega_{\rm c}^2 }{(\beta_{\rm h}\omega_{\rm h}-\beta_{\rm c} \omega_{\rm c})^2} 
     \label{eq:appen_rel_fluc}
.\end{align}
Equation~\eqref{eq:appen_rel_fluc} is rearranged into 
\begin{align}
    l(x,y)\geq 0
\end{align}
where 
\begin{align}
    l(x,y) \equiv &(\coth^2{x} +\coth^2{y} - 2 )(x-y)^2 \nonumber\\
    &- (x^2+y^2)(\coth{x} - \coth{y})^2.
    \label{eq:appen_function_l}
\end{align}

We plot Eq.~\eqref{eq:appen_function_l} in Fig.~\ref{fig:l_poten}.
When the classical Otto cycle is more reliable than the quantum one, $l(x,y)$ is positive. From the figure, it can be seen that as $x$ and $y$ are large enough (high temperature limit), the difference between the quantum and classical cycles decreases. In the cold temperature limit, the difference between the cycles increases, and the quantum cycle's relative error is larger than the classical one. 
Based on this observation, we conjecture Eq.~\eqref{eq:appen_rel_fluc}.

\section{Verification of the enumeration result by the Monte Carlo simulation result}
\label{appendix:B}
\setcounter{figure}{0}  

\begin{figure}[]
 \includegraphics*[width=\columnwidth]{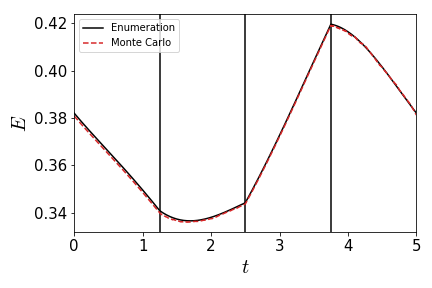}
\caption{The energy trajectory of the working fluid for the classical Otto cycle by the MC simulation result as the red dashed line is compared to that by the enumeration result as the black solid line. To verify the data which violate the conventional TUR, the following parameters are used: $\omega_{\rm h} = 0.724... $, $\omega_{\rm c} = 0.620...$, $T_{\rm h}=0.356...$, $T_{\rm c} = 0.286...$, $m = 1$, $\gamma = 1/4$ and $\tau_{\rm h}=\tau_{\rm c}=\tau_{\rm ch}=\tau_{\rm hc} = 1.25$.}
\label{fig:appen_traj}
\end{figure}

To verify points that violate the conventional thermodynamic uncertainty relation (TUR), we perform Monte Carlo (MC) simulations for a classical nonequilibrium process. In Fig.~\ref{fig:appen_traj}, we plot the energy trajectories of the working fluid for a cyclic steady state. The MC simulation result is shown as the red dashed line and the corresponding enumeration is as the black solid line. It can be seen that the enumeration result is in good agreement with the MC simulation result. 

\begin{figure}[]
 \includegraphics*[width=\columnwidth]{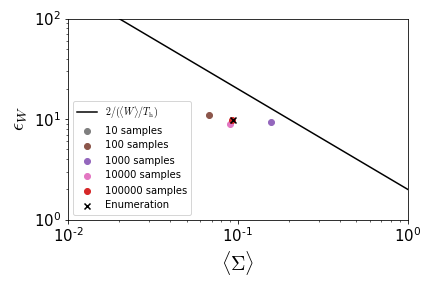}
\caption{The relative work error of the classical Otto cycle is plotted as a function of the entropy production, $\langle \Sigma \rangle$. We use the same parameters as the previous figure. Solid circles ($\bullet$) represent MC results and crosses ($\times$) represent the enumeration results. It can be seen that Monte Carlo results approach to the enumeration result as the number of samples is increased and the point violates the conventional TUR.}
\label{fig:appen_TUR}
\end{figure}

With the same parameters, we plot the relative error of work as a function of entropy production in Fig.~\ref{fig:appen_TUR}. It can be seen that the MC simulation result approaches to  the enumeration result as the number of ensemble is increased, thereby validating the violation of the conventional TUR.

\bibliography{ref-EY11745-LHJ-final}

\end{document}